\documentclass{aastex62}
\usepackage{graphicx}
\graphicspath{{figures/}}
\usepackage{amsmath}
\usepackage{amssymb}
\usepackage{multirow}
\usepackage{enumitem}
\usepackage{color}

\def\approxprop{%
  \def\p{%
    \setbox0=\vbox{\hbox{$\propto$}}%
    \ht0=0.6ex \box0 }%
  \def\s{%
    \vbox{\hbox{$\sim$}}%
  }%
  \mathrel{\raisebox{0.7ex}{%
      \mbox{$\underset{\s}{\p}$}%
    }}%
}

\published{February 2021}
\shorttitle{Ganymede Thermal Properties}
\begin{document}

\title{Ganymede's Surface Properties from Millimeter and Infrared Thermal Emission}

\correspondingauthor{Katherine de Kleer}
\email{dekleer@caltech.edu}

\author[0000-0000-0000-0000]{Katherine de Kleer}
\affil{Division of Geological and Planetary Sciences, California Institute of Technology \\
1200 E California Blvd M/C 150-21 \\
Pasadena, CA 91125, USA}
\author{Bryan Butler}
\affil{National Radio Astronomy Observatory, Socorro, NM 87801}
\author{Imke de Pater}
\affil{Department of Astronomy, UC Berkeley, Berkeley, CA 94720}
\author{Mark A. Gurwell}
\affil{Center for Astrophysics $\vert$ Harvard \& Smithsonian, 60 Garden Street, Cambridge, MA 02138}
\author{Arielle Moullet}
\affil{SOFIA/USRA, Moffett Field, CA 94035}
\author{Samantha Trumbo}
\affil{Division of Geological and Planetary Sciences, California Institute of Technology, Pasadena, CA 91125}
\author{John Spencer}
\affil{Southwest Research Institute, Boulder, CO 80302}

\begin{abstract}
We present thermal observations of Ganymede from the Atacama Large Millimeter Array (ALMA) in 2016-2019 at a spatial resolution of 300-900 km (0.1-0.2'' angular resolution) and frequencies of 97.5, 233, and 343.5 GHz (wavelengths of 3, 1.3, and 0.87 mm); the observations collectively covered all Ganymede longitudes. We determine the global thermophysical properties using a thermal model that considers subsurface emission and depth- and temperature-dependent thermophysical and dielectric properties, in combination with a retrieval algorithm. The data are sensitive to emission from the upper $\sim$0.5 meter of the surface, and we find a millimeter emissivity of 0.75-0.78 and (sub)surface porosities of 10-40\%, corresponding to effective thermal inertias of 400-800 J m$^{-2}$ K$^{-1}$ s$^{-1/2}$. Combined with past infrared results, as well as modeling presented here of a previously-unpublished \textit{Galileo} PPR nighttime infrared observation, the multi-wavelength constraints are consistent with a compaction profile whereby the porosity drops from $\sim$85\% at the surface to 10$^{+30}_{-10}$\% at depth over a compaction length scale of tens of cm. We present maps of temperature residuals from the best-fit global models which indicate localized variations in thermal surface properties at some (but not all) dark terrains and at impact craters, which appear 5-8 K colder than the model. Equatorial regions are warmer than predicted by the model, in particular near the centers of the leading and trailing hemispheres, while the mid-latitudes ($\sim$30-60$^{\circ}$) are generally colder than predicted; these trends are suggestive of an exogenic origin. 
\end{abstract}

\keywords{}

\section{Introduction} \label{sec:intro}
The largest of the Solar System's satellites, Ganymede is a prototypical icy ocean world, hosting a liquid water ocean under a thick ice shell (Kivelson et al. 2002). Its surface is a patchy juxtaposition of dark, ancient, cratered terrain crosscut by bright, icy, tectonized material known as the grooved terrain. The existence of this latter terrain points to a period of significant geological activity in Ganymede's past, which in turn provides clues into its thermal evolution (Schenk et al. 2001). The presence of these two distinct terrain types places Ganymede's surface intermediate in geological history between its neighbors Europa and Callisto: while the grooved terrain shares features with the young, tectonized surface of Europa, the dark terrain resembles Callisto's old, cratered surface. At the same time, Ganymede's intrinsic magnetic field sets it apart from all other Solar System moons, and the interaction of its magnetosphere with the larger jovian magnetosphere in which it orbits leads to complex dynamics and surface modification processes. \par
On Ganymede, the dark terrain characteristics are most consistent with a formation scenario whereby a low albedo component was deposited as Ganymede accreted (Bottke et al. 2013) and was subsequently concentrated on the surface as a lag deposit, or residue left behind after the preferential removal of water ice via processes such as sublimation and impact volatilization (Prockter et al. 2000). However, while detailed spectroscopic data have been used to characterize the distribution of water ice of different grain sizes across Ganymede's surface, the dark material has no spectral features in the near-IR and has not been unambiguously identified by any observations to date (Hansen \& McCord 2004; Ligier et al. 2019). \par
Thermal emission, measured at infrared through radio wavelengths, is sensitive to material properties such as emissivity, thermal conductivity, and heat capacity, the latter two of which are often parameterized through the thermal inertia ($\Gamma$, units of J m$^{-2}$ K$^{-1}$ s$^{-1/2}$ throughout). These properties are in turn determined by the composition, water ice abundance and grain size, surface roughness, and vertical compaction profile, i.e. the density/porosity as a function of depth. Measurements of these properties reveal the relative roles of different processes, both endogenic and exogenic, in determining the structure and evolution of the surface. By constraining near subsurface parameters, measurements of thermal emission from Ganymede in particular can inform models of how ion and electron impacts alter near surface properties of similar icy worlds.\par
Early ground-based eclipse observations in the mid-infrared showed rapid initial cooling of Ganymede upon entering eclipse, from which a very low thermal inertia surface component was inferred (Morrison and Cruikshank 1973). The \textit{Voyager} IRIS instrument measured the 8 to 50 $\mu$m spectrum at many points across Ganymede's surface, detecting temperatures in the 90-150 K range with a strong inverse correlation between temperature and albedo (Hanel et al. 1979; Spencer 1987). A comparison between noon and midnight temperatures for several terrain types in \textit{Galileo} PPR data indicated that a two-component model was sufficient to fit data of most surface regions, with the end-members being dust ($\Gamma=16$) and ice ($\Gamma=1000$) and both dark and grooved terrain lying on a spectrum between the two, in roughly the $\Gamma=70-150$ range (Pappalardo et al. 2004). The bright craters are the only major outliers, appearing much colder than their surroundings during both the night and day (Pappalardo et al. 2004). \par
Observations at longer thermal wavelengths (from sub-mm to $\sim$10 cm) provide complementary information to thermal infrared and near-infrared/optical data: while infrared data are sensitive to the upper $\sim$millimeter of the surface, radio wavelengths sense the upper centimeters to meters and constrain the vertical profile of material properties in this near-surface region where numerous physical and chemical processes are taking place. Such observations have been used for decades to infer the physical properties of the lunar and terrestrial planet surfaces. Tikhonova and Troitskii (1969) derived early constraints on the lunar near-surface density profile based on its radio emission, Rudy et al. (1987) used data from the Very Large Array (VLA) to determine the latitudinal trends in subsurface density of Mars, and Mitchell \& de Pater (1994) used a combination of Berkeley-Illinois-Maryland Association (BIMA) array and VLA data to infer vertical layering within the upper $\sim$meter of Mercury's subsurface. The sensitivity and long baselines of the Atacama Large Millimeter/submillimeter Array (ALMA) now allow us to push this type of investigation to smaller objects farther out in the Solar System, such as the Galilean satellites. ALMA's capabilities for surface characterization have been demonstrated by Trumbo et al. (2017; 2018), who mapped the millimeter emission from Europa with ALMA at 233 GHz and identified regions with unusually high thermal inertia.\par
Early disk-integrated observations of the Galilean satellites at millimeter through centimeter wavelengths revealed differences among the moons (de Pater et al. 1984; Ulich et al. 1984; Altenhoff et al. 1988; Muhleman \& Berge 1991). Ganymede and Europa appear particularly cold at centimeter wavelengths and exhibit correspondingly high radar albedos (Ostro 1982). Muhleman \& Berge (1991) found that Ganymede was the only satellite for which a depressed brightness temperature was measured at a wavelength of 3 mm, although this result is in disagreement with earlier work (Ulich \& Conklin 1976; Ulich 1981) and with the data we present here. \par
We conducted a campaign in 2016-2019 to image the three icy Galilean satellites with ALMA; data were obtained at three frequencies: 97.5, 233, and 343.5 GHz (observations centered at 3, 1.3, and 0.87 mm respectively). These measurements are sensitive to different depth ranges within the upper tens of cm of the surface. Here we present results from the Ganymede observations; analysis of the Europa and Callisto datasets is in progress. Seven observations of Ganymede were obtained with a linear resolution of 300-900 km at Ganymede (0.1-0.2'' angular resolution), collectively covering all longitudes and with a minimum of two observations per frequency band. These data, and the calibration, analysis and imaging procedures, are described in Section \ref{sec:data}. The thermal and radiative transport model and the retrieval procedure used to derive surface properties from the data is described in Section \ref{sec:model}. The thermal maps and best-fit global properties are presented in Section \ref{sec:results}, which also includes the analysis of a previously-unpublished \textit{Galileo} infrared observation for context, and our conclusions are presented in Section \ref{sec:conc}.\par
\section{Data reduction, calibration, and imaging} \label{sec:data}
Data were obtained on seven dates in 2016-2019 with ALMA, located in the Atacama Desert in Chile. Observations were made in three frequency bands: Bands 3, 6, and 7, with central frequencies of 97.5, 233, and 343.5 GHz respectively (wavelengths of 3, 1.3, and 0.87 mm). Two observations of Ganymede were obtained in each frequency band, targeting opposite hemispheres. In addition, a second observation of the leading hemisphere in Band 6 was obtained, offset from the first by 30$^{\circ}$ in longitude. In each band, continuum emission was observed over 4 GHz of total bandwidth, all of which was combined in all analyses. Details of the observations, including the disk-integrated flux densities and brightness temperatures measured on each date, are given in Table \ref{tbl:obs}. Quasars were observed for pointing, bandpass, complex gain, and flux density scale calibration. Quasars 3C279 (J1256-0547) and J1650-2943 were used for flux density scale calibration in 2016-2017 and 2019 respectively. \par
The raw data were reduced and calibrated via the ALMA pipeline; the data were delivered to us in the form of a calibrated Measurement Set (MS). The MS contains, for each target, the amplitude and phase of the cross-correlated signal between each antenna pair. These quantities, referred to as ``visibilities,'' are the fundamental measured quantities of any radio interferometer, and are a sampling of the Fourier transform of the sky brightness temperature distribution at discrete spatial frequencies. Processing of the MS is performed using the Common Astronomy Software Applications (CASA) package (McMullin et al. 2007), and the MS containing each observation is processed and imaged separately. We perform an iterative imaging and self-calibration procedure (Cornwell \& Fomalont 1999) on the visibilities to improve the phase calibration and produce an image that can be compared with models. For the first iteration, self-calibration is performed using a model of a limb-darkened disk the size and brightness temperature of Ganymede (Table \ref{tbl:obs}), where the modeled limb darkening is described by $\text{cos}^{0.2}(\theta)$ for emission angle $\theta$. The choice of limb darkening parameter has negligible effect on the final image since the model is only used for the first iteration of self-calibration. The calibrated visibilities are then imaged, and the resultant image is used as the model for the next iteration of self-calibration. Self-calibration is performed on the visibility phases only, with a progressively shorter solution interval for each iteration down to a minimum solution interval of 8 seconds; shorter solution intervals do not result in further improvements and are therefore not employed. Imaging is performed with the CASA task \textbf{tclean} using a multi-scale multi-frequency algorithm for deconvolution (Rau \& Cornwell 2011) assuming that intensity is linear in frequency over our frequency interval and fitting for the slope at each spatial pixel. We use a robust parameter of 0, which is a good compromise between signal-to-noise and resolution.\par
We find the disk-integrated flux density by fitting the visibilities with a uniform disk model (CASA's \textbf{uvmodelfit} task) to find the zero-spacing value. The visibilities are used rather than the image because the image is derived from a Fourier transform of the visibilities with subsequent deconvolution, which can introduce artifacts, while the visibilities are the direct measured quantity. As a check on the accuracy of the flux densities in the final images, we also extract the disk-integrated flux density from each image via aperture photometry, and find it to be within 1-2\% of the flux density derived from the visibility fitting. We then use the fitted flux density to derive the disk-averaged brightness temperature $T_b$, using the known geometry at the time of observation (Table \ref{tbl:obs}), by inverting the equation:
\begin{equation}
F_{\nu}=10^{26}\frac{\pi R_G^2}{4.25\times10^{10}} \frac{2h\nu^3}{c^2}\Big[\frac{1}{e^{h\nu/k_bT_b}-1}-\frac{1}{e^{h\nu/k_bT_{\text{cmb}}}-1}\Big]
\end{equation}
where $R_G$ is the radius of Ganymede in arcseconds, $F_{\nu}$ is in Jansky units, and $T_{\text{cmb}}=2.725$ K. The quantity $4.25\times10^{10}$ is the number of square arcseconds in one steradian. The set of final self-calibrated images from all Ganymede observations is shown in Figure \ref{fig:data}.\par
\subsection{Flux density scale calibration} \label{sec:fluxcal}
ALMA performs a nominal flux density scale calibration on delivered datasets based on the measured flux densities of specific calibrators at the time of observation. However, the flux density and spectral shape of the calibrators are variable, and the assumed flux density of a given calibrator in a given band and date is calculated from a model based on observations at different frequencies and times. We therefore check the flux density scale calibration as follows.\par
The flux density of the flux density scale calibrator is retrieved from the ALMA calibrator catalogue for all dates and frequencies at which its flux density was measured; typically measurements were made in Bands 3 and 7 at an interval of roughly two weeks. For each date when measurements were made in both Bands 3 and 7, we derived the exponent $\alpha$ in the intensity scaling $I(\nu)\propto\nu^{\alpha}$ appropriate for quasars. The fit parameters are then used to calculate the flux densities calculated at other frequencies. The fit uses the uncertainties on the measurements to derive uncertainties on the line fit parameters, which are propagated into uncertainties on the flux density at other frequencies. The exponent $\alpha$ is typically found to lie between -0.8 and -0.5, which matches the expected value for quasars (Ennis et al. 1982). \par
For a given frequency of observation, the flux density of the calibrator on all dates on which it was measured are calculated in this way from the measurement frequencies. The calibrator timeline for a given frequency is then used to predict the calibrator flux densities on the date of Ganymede data acquisition, which is typically not a date on which the calibrator flux density was measured relative to other flux density scale standards. The calibrator flux density on the observing date is estimated from a fit to the nearby measurements, with an error estimate derived from the quality of the fit and the original measurement uncertainties. We find that the calibrator flux densities estimated by our methods differ from the ALMA pipeline calibration by more than 10\% in some cases, in an unsystematic way. Similar deviations were also found by Trumbo et al. (2018). We apply a gain factor based on the estimated flux density of the calibrator to all the data in the MS, thereby placing that data on an absolute flux density scale.\par
Even after this correction, the accuracy of ALMA calibration is no better than 5\% due to other known issues in how the calibration is applied, based on our experience with several past ALMA datasets. We therefore increase the flux density scale calibration uncertainties to no less than 5\% in all datasets.\par
The Band 6 dataset obtained in 2019 has a lower brightness temperature than all other datasets at all frequencies, even after the correction described above, and is about 8\% lower in brightness temperature than the other Band 6 measurement that overlaps it in longitude. The 2019 data were obtained 2-3 years later than the rest of the observations, and used a different flux density scale calibrator from the rest of the data. The difference cannot be explained by the changing solar distance, since Ganymede was closer to the sun in 2019 than in 2016-2017, nor by Jupiter proximity, since this dataset was one of three taken with Ganymede at a Jupiter distance of 120-140'' (the remainder had a Jupiter distance of $>$200''), and Galilean moon observations have been successfully calibrated much closer to Jupiter (e.g. de Pater et al. 2020). The solar phase angle was higher by 8.1$^{\circ}$ in 2019 compared to the dataset that overlaps the 2019 data in longitude within the same frequency band, which could result in a suppressed brightness temperature due to viewing incompletely-illuminated particles on a rough surface. However, Francis et al. (2020) show that depending on the frequency band and calibrator behavior the flux density scale calibration from ALMA may be even poorer than 10\%. The results presented here therefore do not include this dataset when presenting global brightness temperatures and emissivities, but do include it when showing residual images and maps because all fits presented here allow for a different emissivity value for each dataset, which absorbs the calibration error.
\begin{table}[h!]
\begin{center}
\caption{Observations and Disk-Integrated Brightness Temperatures\label{tbl:obs}}
\begin{tabular}{c|c|c|c|c|c|c|c|c|c|c}
Date & Time & Ang Diam & Beam & Ob Lat & Ob Lon & Phase Ang & Freq & $\lambda$ & F$_{\nu}$ & T$_b$ \\
{[}UT] & [UT] & [''] & [''] & [$^{\circ}$N] & [$^{\circ}$W] & [$^{\circ}$] & [GHz] & [mm] & [Jy] & [K] \\
\hline
2017-09-06 & 15:18 & 1.17 & 0.1$\times$0.21 & -2.6 & 256 & 6.7 & 91.5 & 3.3 & 0.57$\pm$0.03 & 90.6$\pm$4.4 \\
 & & & & & & & 103.5 & 2.9 & 0.73$\pm$0.04 & 90.4$\pm$4.4 \\
\hline
2017-11-14 & 12:53 & 1.14 & 0.13$\times$0.15 & -2.9 & 108 & 2.6 & 91.5 & 3.3 & 0.53$\pm$0.03 & 89.9$\pm$4.3 \\
 & & & & & & & 103.5 & 2.9 & 0.68$\pm$0.03 & 90.2$\pm$4.3 \\
\hline
2016-10-13 & 13:57 & 1.13 & 0.15$\times$0.2 & -2.1 & 325 & 2.4 & 225.0 & 1.3 & 3.43$\pm$0.17 & 99.3$\pm$4.7 \\
 & & & & & & & 241.0 & 1.2 & 3.96$\pm$0.2 & 100.1$\pm$4.7 \\
\hline
2016-10-15 & 13:19 & 1.13 & 0.15$\times$0.22 & -2.1 & 63 & 2.7 & 225.0 & 1.3 & 3.1$\pm$0.15 & 89.8$\pm$4.2 \\
 & & & & & & & 241.0 & 1.2 & 3.6$\pm$0.18 & 91.3$\pm$4.3 \\
\hline
2019-08-26 & 19:07 & 1.46 & 0.09$\times$0.18 & -2.3 & 31 & 10.8 & 225.0 & 1.3 & 4.75$\pm$0.24 & 83.6$\pm$3.9 \\
 & & & & & & & 241.0 & 1.2 & 5.46$\pm$0.27 & 84.1$\pm$3.9 \\
\hline
2016-10-24 & 11:50 & 1.14 & 0.11$\times$0.22 & -2.2 & 151 & 3.9 & 337.5 & 0.89 & 7.62$\pm$0.4 & 98.5$\pm$4.8 \\
 & & & & & & & 349.5 & 0.86 & 8.29$\pm$0.45 & 100.1$\pm$5.0 \\
\hline
2016-10-26 & 12:08 & 1.14 & 0.15$\times$0.24 & -2.2 & 252 & 4.2 & 337.5 & 0.89 & 7.62$\pm$0.49 & 98.2$\pm$5.8 \\
 & & & & & & & 349.5 & 0.86 & 8.1$\pm$0.53 & 97.8$\pm$5.9 \\
\hline
\end{tabular}
\end{center}
\end{table}
\section{Thermal model} \label{sec:model}
After calibration and imaging, each ALMA dataset consists of a map of brightness temperature across Ganymede's disk at the orientation and viewing geometry at the time of observation, where brightness temperature $T_b$ is defined as the temperature of a blackbody that produces the observed intensity $I_{\nu}$ at a frequency $\nu$:
\begin{equation}\label{eqn:tb}
T_b=\frac{h\nu}{k_b}\frac{1}{\textrm{ln}\Big(1+\frac{2h\nu^3}{c^2(I_{\nu}+I_{\nu,\text{cmb}})}\Big)}.
\end{equation}
\indent We fit each ALMA image with a thermal model, which treats transport of heat by conduction and radiation within a solid or porous surface. The model constructs a temperature profile for every latitude and longitude on Ganymede's surface, fixing albedo from past spacecraft observations and incorporating the dependence of thermal properties on porosity, grain size, and temperature. The thermal emission is integrated along the line-of-sight in the subsurface, as described in Section \ref{sec:radtrans}, according to the dielectric properties, which likewise depend on composition, porosity, and temperature as described in Section \ref{sec:dielectricprop}. Cold, icy surfaces are highly transparent to millimeter radiation, which may arise from depths of tens of cm in the surface where the temperature may differ from that of the surface by $>$10 K in a way that varies systematically across the surface and between terrains. Interpretation of millimeter observations of such surfaces, in contrast to infrared observations, therefore requires accounting for subsurface emission.\par
The free parameters in the fits are the millimeter emissivity and the porosity of the material (or surface porosity when a depth-dependent porosity is used). Both the thermophysical and dielectric properties depend strongly on temperature and hence on time of day and depth. Because the thermal inertia is not a constant value in time, spatial coordinate, or depth, the thermal inertia reported throughout is an effective thermal inertia $\Gamma_{\text{eff}}$ calculated from the porosity as described in Section \ref{sec:radtrans}. In our model, varying the porosity results in changes to the thermal conductivity, the density, and the dielectric properties, so that as porosity is varied both the thermal and the radiative transport components of the model vary in a self-consistent way. \par
The grain size and the ice fraction can be varied but are not free parameters in the fits; the effect on our results of different parameter choices will be described in Section \ref{sec:thermalprop}, which deals with the treatment of the thermal properties. The heat transport model is described in Section \ref{sec:therm}, and Sections \ref{sec:albedomap} and \ref{sec:retrievals} describe the creation of an albedo map for the model, and the retrieval algorithm.\par
\subsection{Radiative transport}\label{sec:radtrans}
Thermal emission is integrated through the subsurface along the viewing path:
\begin{equation} \label{eqn:Ieff}
I_{\nu} = \frac{\int_{0}^{\infty}I_{\nu}(z)e^{-z/(\delta_{elec,\nu}(T(z))cos\theta_t)}dz}{\int_{0}^{\infty}e^{-z/(\delta_{elec,\nu}(T(z))cos\theta_t)}dz}
\end{equation}
where $z$ is the depth coordinate, $I_{\nu}(z)$ is calculated from $T(z)$ via the full Planck function, $T(z)$ is the output of the numerical thermal model (Section \ref{sec:therm}), and 
\begin{equation}
cos\theta_t=\sqrt{1-\frac{sin^2\theta}{\epsilon'}}
\end{equation}
for emission angle $\theta$ and real part of the dielectric constant $\epsilon'$, which is a function of porosity and composition only. The electrical skin depth $\delta_{elec}$, also known as the penetration depth, describes the depth over which an electromagnetic wave passing through a material is attenuated by a factor of 1/$e$ and is described in Section \ref{sec:dielectricprop}. \par
The spectral emissivity $E_{\nu}$ is defined such that
\begin{equation}
I_{\nu,obs} = E_{\nu}I_{\nu,model}
\end{equation}
and is a free parameter in the fits. An effective temperature $T_{\text{eff}}$ and effective thermal inertia $\Gamma_{\text{eff}}$ are similarly calculated following Equation \ref{eqn:Ieff}:
\begin{equation}
T_{\text{eff}} = \frac{\int_{0}^{\infty}T(z)e^{-z/(\delta_{elec,\nu}(T(z))cos\theta_t)}dz}{\int_{0}^{\infty}e^{-z/(\delta_{elec,\nu}(T(z))cos\theta_t)}dz}
\end{equation}
\begin{equation}
\Gamma_{\text{eff}} = \frac{\int_{0}^{\infty}\Gamma(z)e^{-z/(\delta_{elec,\nu}(T(z))cos\theta_t)}dz}{\int_{0}^{\infty}e^{-z/(\delta_{elec,\nu}(T(z))cos\theta_t)}dz}
\end{equation}
where $\Gamma(z)$ is calculated from the thermal properties at each time and depth coordinate as described in Section \ref{sec:thermalprop}. $I_{\nu,obs}$ is the quantity that is directly compared with the observations, and $\Gamma_{\text{eff}}$ is the value reported in the results as effective thermal inertia. \par
\subsection{Dielectric properties} \label{sec:dielectricprop}
The sensitivity of a given wavelength of observation $\lambda$ to thermal emission from a depth $z$ within the subsurface is set by the dielectric properties of the material, in particular the electrical skin depth $\delta_{elec}$:
\begin{equation}
\delta_{elec}=\frac{\lambda}{4\pi \kappa}
\end{equation}
where $\kappa$ is the imaginary part of the complex refractive index $\tilde{n}=n+i\kappa$. Formalisms calibrated to laboratory data provide estimates of the real and imaginary part of the dielectric constant $\tilde{\epsilon}=\epsilon'+i\epsilon''$ as a function of properties such as frequency, temperature, porosity, and composition. The loss tangent $\epsilon''/\epsilon'$ describes the degree of dissipation of electrical energy from passing through the medium and is inversely related to the transparency of the medium to radiation. The complex refractive index is related to the dielectric constant through 
\begin{equation}
\tilde{\epsilon} = \tilde{n}^2.
\label{eqn:eps}
\end{equation}
Equating the real part of each side of Equation \ref{eqn:eps} yields
\begin{equation}
\epsilon'=n^2-\kappa^2
\end{equation}
from which $\kappa$ can be expressed in terms of the experimentally-determined dielectric properties:
\begin{equation}
\kappa(p,f_{\text{dust}},\nu,T)=\sqrt{\frac{1}{2}\Big(|\tilde{\epsilon}(p,f_{\text{dust}},\nu,T)|-\epsilon'(p,f_{\text{dust}})\Big)}
\label{eqn:kappa}
\end{equation}
where $|\tilde{\epsilon}(p,f_{\text{dust}},\nu,T)|$ is the complex modulus and our model treats the dielectric properties as dependent on porosity $p$, dust mass fraction $f_{\text{dust}}$, frequency $\nu$, and temperature $T$ as indicated in Equation \ref{eqn:kappa}. The electrical skin depth calculated in this way enters directly into the radiative transport model described in Section \ref{sec:radtrans}. \par
Ganymede's surface is composed of ice, likely porous at the depths accessed by our observations, combined with dark materials that have not been definitively identified but likely include CO$_2$, SO$_2$, organics, and sulfur-bearing species (McCord et al. 1997). The dielectric properties of ice and snow have been measured in the laboratory across a range of temperatures, frequencies, and porosities. However, the dielectric properties of dirty ice mixtures are less well studied, particularly over the frequencies, compositions, and temperatures of interest for our observations. We therefore adopt the dielectric properties of a porous dust/ice mixture. We use the formalism of Hufford (1991) for the frequency and temperature dependence of the dielectric properties of pure ice in the 100-200 K and 100-350 GHz ranges; this formalism is calibrated to experimental data, of which the most relevant for our purposes is the 1 mm data from Mishima et al. (1983), and is extended to higher frequencies by Jiang \& Wu (2004). We then incorporate the porosity of the ice following the formalism from Tiuri et al. (1984). We assume a dust volume fraction of 0.2 and use the properties of meteoritic dust following Heggy et al. (2012), combining the properties of the two materials using the Maxwell-Garnett approximation (Choy 1999).  In the resultant model, $\epsilon'$ is a function of porosity and dust mass fraction (Brouet et al. 2016), while $\epsilon''$ is a function of porosity, dust mass fraction, temperature, and frequency. A better understanding of the composition of Ganymede's dark surface materials, combined with future laboratory data at 100-350 GHz covering the compositions and temperatures appropriate for Ganymede's surface, would greatly improve the accuracy of the dielectric model. \par
\begin{figure}[ht]
\centering
\includegraphics[width=14cm]{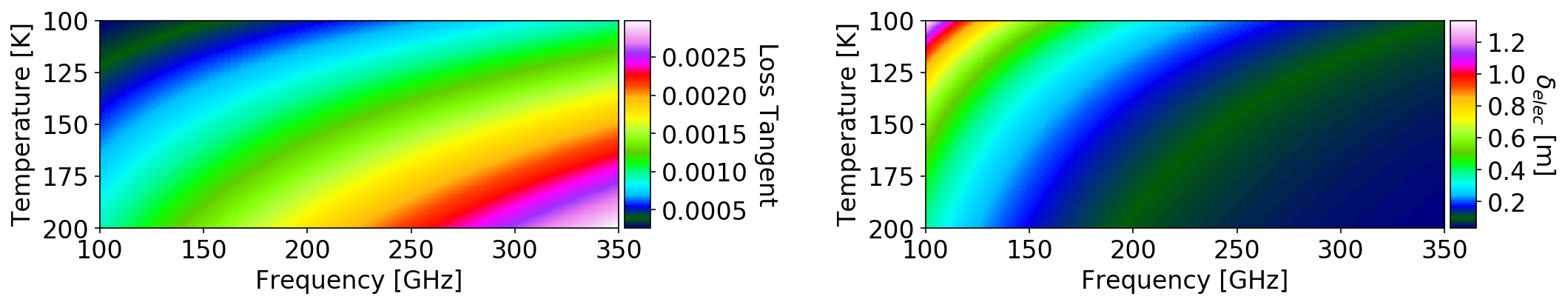}
\caption{The loss tangent and corresponding electrical skin depth of millimeter waves as a function of temperature and frequency, in the frequency range of ALMA and temperatures of Ganymede's surface, for a surface with 50\% porosity. \label{fig:dp}}
\end{figure}

The resultant values for the loss tangent and electrical skin depth over the relevant domains are shown in Figure \ref{fig:dp}; the losses are lowest and electrical skin depths largest for low-frequency radiation at low temperatures. For 100 GHz emission from a 50\% porosity surface near 100 K, $\delta_{elec}$ may be as deep as a meter, while at 350 GHz and 200 K ice, as shallow as 3 cm. At 150 K, this model yields $\delta_{elec}=$ 0.58 m, 0.11 m, and 0.05 m at frequencies of 97.5, 233, and 343.5 GHz ($\lambda=$3, 1.3, 0.87 mm) respectively. The temperature dependence of porous ice transparency is sufficiently steep at millimeter wavelengths that spatial variations in electrical skin depth arising from differences in albedo and latitude can result in brightness temperature variations of several K between different surface regions. In particular, high-albedo features such as impact craters appear colder not just due to albedo, but also due to the fact that the observed emission arises from greater depths where temperatures are lower during the daytime. \par
\subsection{Thermal properties} \label{sec:thermalprop}
The thermal model includes cases for both a solid and porous surface; the former case treats heat flow in the subsurface via conduction only, while the latter treats conductivity as a combination of the radiative conductivity through pores and the contact conductivity at grain contacts. In the solid surface case, the effective thermal conductivity $k_{\text{eff}}$ is equal to the thermal conductivity of solid ice $k_{s}$, which depends only on temperature (Klinger 1980) and is given by 
\begin{equation}
k_s(T) = \frac{567}{T}.
\end{equation}
\indent In the porous medium case,
\begin{equation}
\label{eqn:Keff}
k_{\text{eff}}=k_r + k_c,
\end{equation}
for a radiative term $k_r$ and a contact term $k_c$. The radiative term is
\begin{equation}
\label{eqn:Kr}
k_r=BT^3
\end{equation}
where
\begin{equation}
B = 8RF_E\sigma,
\end{equation}
and $R$ is the grain size, $\sigma$ is the Stefan-Boltzmann constant, and 
\begin{equation}
F_E = \frac{1.34p}{1-p}
\end{equation}
for porosity $p$, following Ferrari \& Lucas (2016) and the model of Gundlach \& Blum (2012) and assuming an ice grain emissivity of 1. The parameter $B$ is roughly $10^{-11}-10^{-9}$ W/m/K$^4$ for 1 mm grains over a range of porosities, decreasing with decreasing grain size to $10^{-14}-10^{-12}$ W/m/K$^4$ for 1 $\mu$m grains. For comparison, a value of $4\times10^{-11}$ has been used for the Moon (Hayne et al. 2017).\par
\begin{figure}[ht]
\centering
\includegraphics[width=14cm]{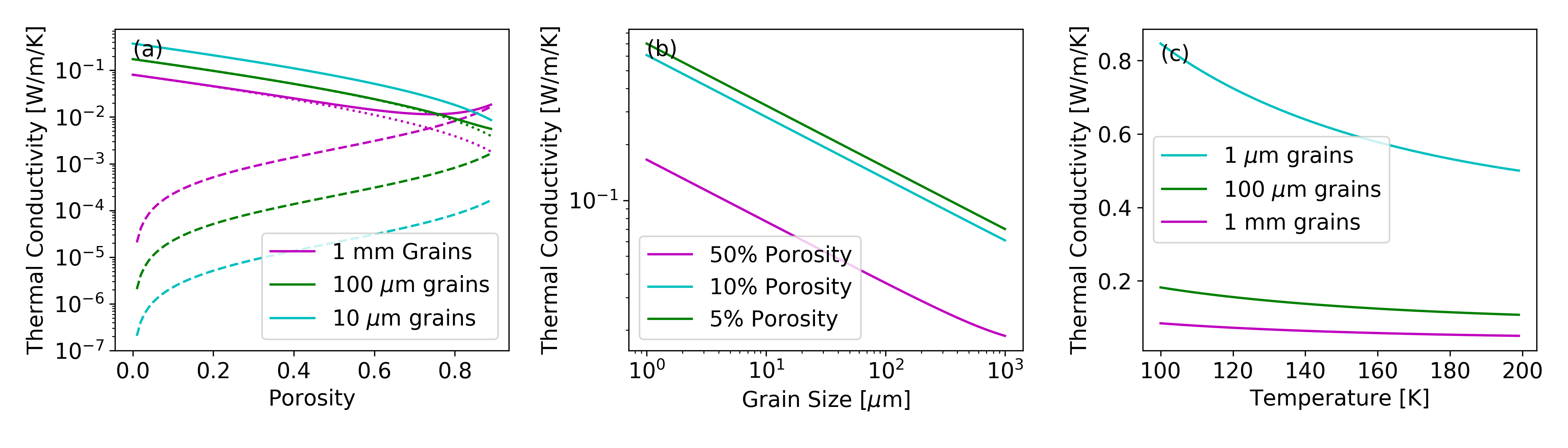}
\caption{Thermal conductivity of a porous icy surface as a function of different surface properties: (a) Porosity, for three grain sizes representative of Ganymede's surface and a temperature of 150 K. The dashed lines are the radiative contribution, the dotted lines the conductive contribution, and the solid lines the total effective conductivity; (b) Grain size, for three porosities and a fixed temperature of 150 K; (c) Temperature, for a fixed porosity of 10\% and three grain sizes. \label{fig:kcond}}
\end{figure}

For the heat conducted through grain contacts, we follow Ferrari and Lucas (2016) in using the model of Johnson et al. (1971) and Gusarov et al. (2003) for tight grain contacts, noting that models for highly porous media and loose grain contacts cannot achieve the high thermal conductivities required to fit observations of Ganymede at millimeter wavelengths. The contact conductivity $k_c(p,R,T)$ is the solid material conductivity $k_s(T)$, reduced by a factor $\phi(p)$ that acts to decrease the bulk conductivity to account for porosity, and multiplied by the Hertz factor $h(T,R)$ which represents the ratio of the grain contact area (or neck size) to total grain size:
\begin{equation}
\begin{aligned}
\label{eqn:Kc}
k_c(p,R,T) &= h(T,R)\times \phi(p) \times k_s(T) \\
&= \Big(\frac{9}{4} \frac{(1-\nu^2)\gamma \pi}{E(T)R}\Big)^{1/3}\times \frac{(1-p)n_c(p)}{\pi} \times k_s(T),
\end{aligned}
\end{equation}
where $\nu=0.33$, $\gamma=0.37$ J/m$^2$, Young's modulus $E(T)=6.6\times10^{9}(4.276-0.012T)$ N/m$^2$, the characteristic number of contacts per grain $n_c(p)=2.17e^{0.0019\rho_{s}(1-p)}$, and $\rho_s=934$ kg/m$^3$ is the density of solid ice. The reader is referred to the references above for the determination of these parameterizations from experimental data. Figure \ref{fig:kcond} shows the effective thermal conductivity as a function of porosity, grain size, and temperature.\par
The heat capacity of water ice is given by $c_p(T)=7.49T+90$ J/K (Klinger 1980) and is similar between water ice in its crystalline and amorphous forms. This simple formulation matches experimental data well down to roughly 100 K though it diverges at lower temperatures (Shulman 2004).\par
The thermal inertia of a material is defined as
\begin{equation}
\label{eqn:TI} 
\Gamma=\sqrt{k_{\text{eff}}(p,R,T)\rho_{\text{eff}}(p) c_p(T)}.
\end{equation} 
\indent This parameter sets the time-of-day temperature variations: a low thermal inertia results in a high amplitude diurnal temperature cycle and strong day-to-night temperature contrast, while a high thermal inertia results in a flatter diurnal temperature cycle and a small temperature contrast between night and day. For a given material composition, $\Gamma$ is a function of porosity $p$, grain size $R$, and temperature $T$ and therefore varies with surface location and depth. However, when the radiative conductivity is negligible, the thermal inertia has only a minor temperature dependence because $c_p\approxprop T$ while $k_s \propto T^{-1}$. The dependence of thermal inertia on porosity and grain size is shown in Figure \ref{fig:InertiaModel}. \par
Grain sizes of 1 $\mu$m - 10 cm were used in testing, but the results presented in this paper are all calculated for a fixed grain size of $R=1$ $\mu$m. This choice was made despite the fact that particle sizes in the 50 $\mu$m - 1 mm range have been inferred from observations (Ligier et al. 2019; Stephan et al. 2020) because models with larger grain sizes provide much poorer fits to the data. This is because such models have a limited neck-to-diameter ratio  (i.e. contact area to grain size) even in the tight contact case adopted here, and hence do not achieve sufficiently high thermal inertias because high conductivities are only achieved at low densities, when radiation dominates the heat transport. A conductivity model that incorporates grain neck growth due to sintering could perhaps reconcile the high thermal conductivities inferred from our data with the larger grain sizes inferred from optical-infrared data. Note, however, that the highest conductivity case, that of solid ice, provides a much poorer fit to the data than the porous models.\par
\subsection{Temperature profile and thermal transport} \label{sec:therm}
\begin{figure}[ht]
\centering
\includegraphics[width=14cm]{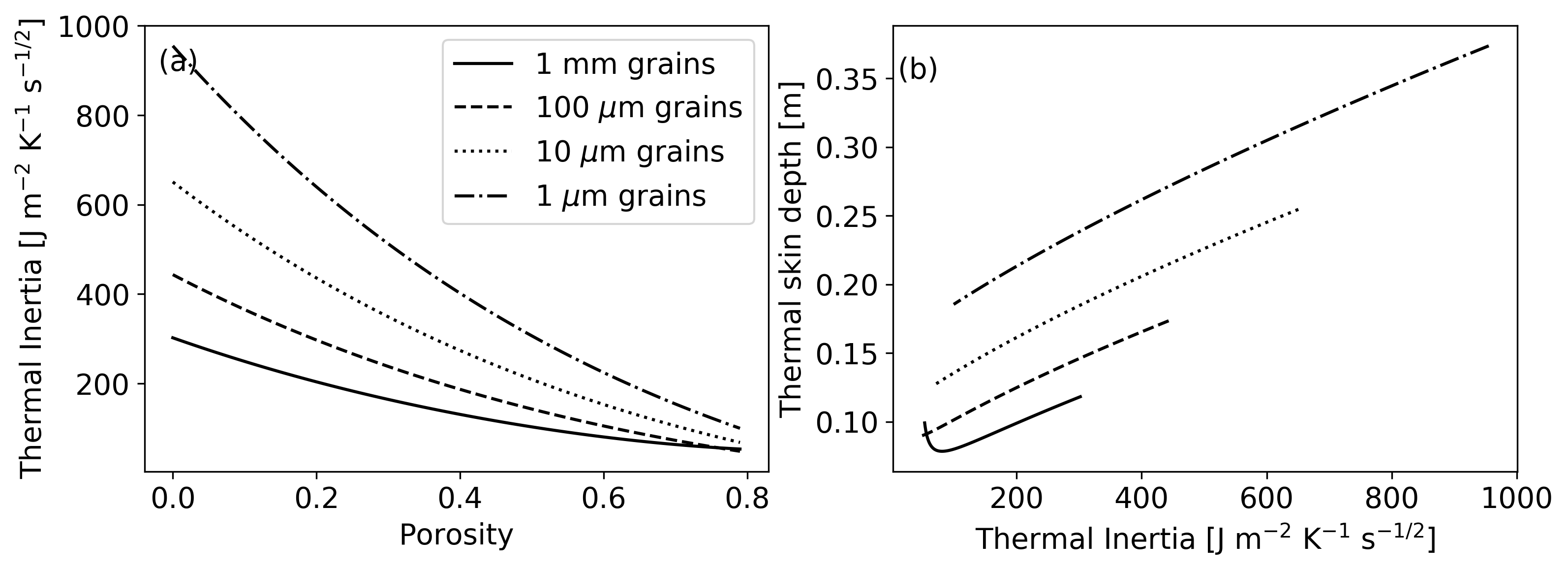}
\caption{(a) Model thermal inertia as a function of porosity for different grain sizes. The thermal inertia is nearly constant in temperature over the temperatures relevant to the Galilean satellites. (b) The relationship between thermal skin depth and thermal inertia. \label{fig:InertiaModel}}
\end{figure}

The thermal properties and their parameter dependencies enter into the model through the numerical treatment of thermal transport, which is as follows and builds on a large volume of work over many decades (e.g. Morrison 1969; Spencer et al. 1989; Mitchell \& de Pater 1994). A three-dimensional temperature grid is set up for Ganymede's (sub)surface with a spatial resolution of 5$^{\circ}$ in latitude and longitude. The vertical layer thickness is $\delta_{therm}$/m at the surface and increases geometrically by a factor of (1$+$1/n) with each layer to match the local resolution of the model to the degree of temperature variation; values of m=10 and n=5 were used for the model runs, and the full vertical extent of the modeled region is always a minimum of 10$\delta_{therm}$. The diurnal thermal skin depth $\delta_{therm}$ is defined as the vertical distance over which the amplitude of the diurnal thermal wave is attenuated by a factor of $1/e$ and is given by 
\begin{equation}
\label{eqn:dtherm1}
\delta_{therm}=\sqrt{\frac{k_{\text{eff}}(p,R,T_{\text{eff}})P}{\pi \rho_{\text{eff}}(p) c_p(T_{\text{eff}})}}
\end{equation}
where $P$ is the diurnal period, $\rho_{\text{eff}}(p)=\rho_{s}\times(1-p)$ and other quantities are defined above. $\delta_{therm}$ is related to the effective thermal inertia $\Gamma_{\text{eff}}$ by:
\begin{equation}
\label{eqn:dtherm}
\delta_{therm}=\sqrt{\frac{P}{\pi}}\frac{k_{\text{eff}}(p,R,T_{\text{eff}})}{\Gamma_{\text{eff}}}.
\end{equation}
\indent Although $\Gamma_{\text{eff}}$ appears in the denominator in Equation \ref{eqn:dtherm}, the thermal skin depth is higher for higher thermal inertias due to the strong dependence of thermal conductivity on density (see Figure \ref{fig:InertiaModel}). \par
For grain sizes of 1 $\mu$m - 1 mm, temperatures of 100-200 K, and 5-50\% porosity, $\delta_{therm}$ is in the 5-50 cm range (see Figure \ref{fig:InertiaModel}), compared to 0.6-1.0 meters for solid ice. The solid ice case corresponds to a thermal inertia of $\sim$2000, which is inconsistent with observations of any icy Solar System body at any wavelength and provides a poorer fit to our data than the porous case; the results presented here therefore use the conductivity model for the porous medium case. \par
Given the parameter values described above, at ALMA Band 3 the emission arises from beneath a thermal skin depth under nearly all combinations of parameters, and may arise from several thermal skin depths down in the surface under certain conditions. In Bands 6 and 7, the penetration depth is 0.2-0.5$\delta_{therm}$. The temperature at the depths from which emission is arising therefore differs from the surface temperature by a few K up to $>$10 K in some cases, which necessitates including emission from the subsurface even for the shortest (0.87 mm) ALMA wavelengths in our dataset.\par
The temperature profile at each spatial location is initialized as a simple exponential decrease between some initial boundary conditions for the surface temperature at the lower and upper boundary of the model:
\begin{equation}
T_0(z) = (T_{surf}-T_{deep})e^{-z/\delta_{therm}}+T_{deep}
\end{equation}
where $T_i(z)$ is the temperature at depth $z$ computed at the $i$th timestep. $T_{deep}$ is the temperature at a depth of 10$\delta_{therm}$ and is assumed to be constant over a diurnal cycle; it is initialized to the temperature that corresponds to the diurnally-averaged radiation assuming a surface in instantaneous equilibrium with incident sunlight, and is evolved until model convergence as described later in this section. After initialization, the temperature profile is then evolved with time according to the 1-D diffusion equation
\begin{equation}
{\rho_{\text{eff}} c_p}\frac{\partial T}{\partial t} =\frac{\partial}{\partial z}\Big(k_{\text{eff}}\frac{\partial T}{\partial z}\Big)
\label{eqn:heat}
\end{equation}
where the constants and variables are as described above. Given that $k$ is a function of $\rho$ and $T$, which themselves vary with $z$, Equation \ref{eqn:heat} becomes
\begin{equation}
\rho_{\text{eff}} c_p \frac{\partial T}{\partial t} = \frac{\partial k_{\text{eff}}}{\partial T} \Big(\frac{\partial T}{\partial z}\Big)^2 + \frac{\partial k_{\text{eff}}}{\partial \rho_{\text{eff}}} \frac{\partial \rho_{\text{eff}}}{\partial z} \frac{\partial T}{\partial z} + k_{\text{eff}} \frac{\partial^2 T}{\partial z^2}
\end{equation}
which is implemented in the finite-difference approximation. The time-evolution of the system then amounts to adjusting the new temperature of each layer based on the balance between heat exchange through the upper and lower boundaries as set by the temperature differences between the layers.\par
The porosity of the near-surface likely varies with depth, but the nature of such variations are only poorly constrained by data to date. As such, the main results presented in this paper fit the data at each frequency independently of one another, with porosity as a free parameter and constant with depth. However, models with depth-dependent porosities are also tested based on the results of the initial fits, and will be shown in Section \ref{sec:results}. In these models we parameterize $\rho_{\text{eff}}$ by an exponential decrease in porosity with depth:\par
\begin{equation}
\rho_{\text{eff}}(z) = \rho_s(1-p_{\text{surf}} e^{-z/\delta_{\text{compact}}})
\label{eqn:rho_vs_z}
\end{equation}
where $p_{\text{surf}}$ is the porosity at the surface and $\delta_{\text{compact}}$ is a specified compaction length scale.\par
The temperature at the surface is calculated by evaluating Equation \ref{eqn:heat} at $z=0$ and imposing the boundary condition
\begin{equation}
k_{\text{eff}}\frac{\partial T(z,t)}{\partial z}\Bigr|_{z=0} = (\epsilon \sigma T^4(0,t)-F_{solar})
\end{equation}
where $\epsilon$ is the bolometric emissivity. $F_{solar}$ is the solar energy absorbed by a given surface area unit at a given time point, as set by the albedo of that surface region, the incidence angle, the distance from the Sun to Ganymede at the time of observation, and the absence of sunlight during eclipse by Jupiter. The model treats sunlight absorption in just the upper layer of the model. The surface temperature is thus determined by three sources and sinks: heat exchange with the layer below, energy radiated into space, and absorbed sunlight. The albedo is fixed based on measurements from past spacecraft missions; the albedo map used in the model is described in Section \ref{sec:albedomap}.\par
The model is initialized at every latitude and longitude point and evolved in $N_t$ timesteps per diurnal period $P$, where $N_t$ is set to the smallest number that preserves numerical stability given the layer thickness, typically $\sim$1000. At every timestep, the surface boundary condition is solved iteratively using Newton's method (see e.g. Hayne et al. 2017), the time-averaged balance of absorbed and emitted radiation is calculated, and the temperature at the lower boundary is adjusted; the model is considered converged when the maximum change between subsequent cycles for any individual timestep is less than 0.1 K for every spatial location. Convergence is typically achieved within $\sim$5 diurnal cycles. The model assumes thermal balance at every location over each diurnal cycle and does not treat seasonal periodic temperature changes in the subsurface beyond accounting for the distance of Ganymede from the Sun at the time of each observation when computing solar insolation. \par
\begin{figure}[ht]
\centering
\includegraphics[width=14cm]{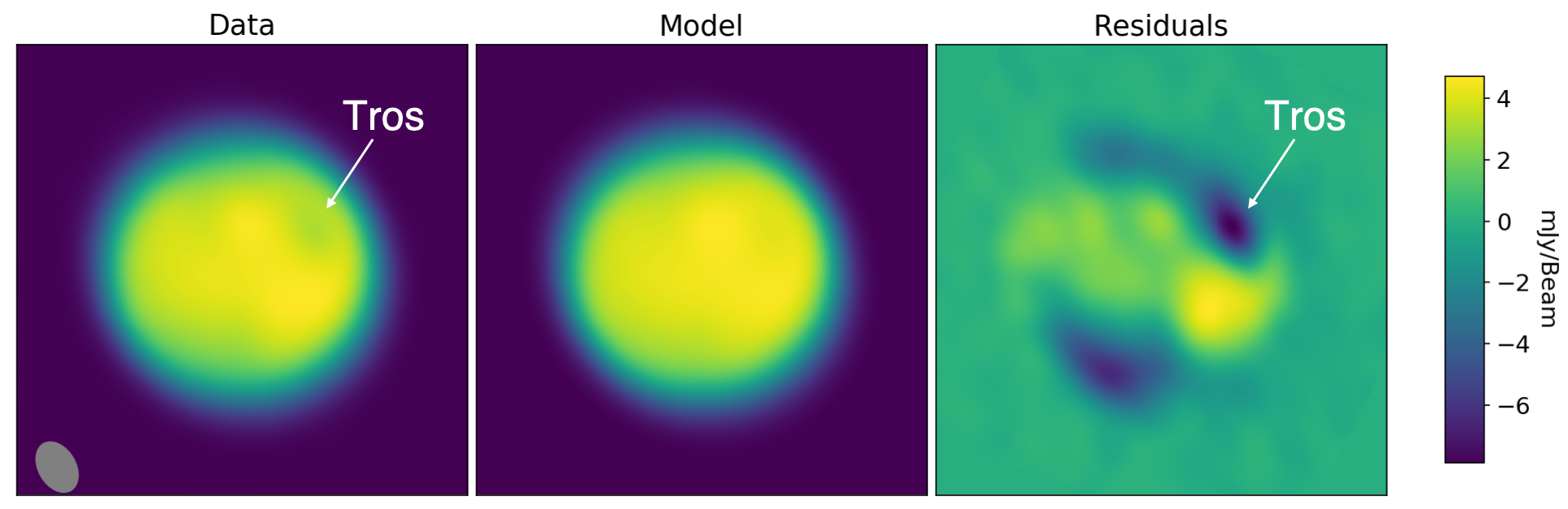}
\caption{Example data, model, and residuals (data-model) for Ganymede at 233 GHz (1.3 mm) for a porosity of 35\%, corresponding to an effective thermal inertia of 450. The residual surface brightness corresponds to 5-10 K in brightness temperature and is a factor of 10 above the noise level for this observation. The impact crater Tros is labeled and appears anomalously cold. \label{fig:demomodel}}
\end{figure}

The output of the model, for a given set of input parameters, is the temperature at every spatial and vertical location on the 3D grid, at all $N_t$ time points during a diurnal period. This temperature grid is then used to calculate the brightness temperature distribution across Ganymede's disk, as described in Section \ref{sec:radtrans}. Figure \ref{fig:demomodel} shows as a demonstration a single observation alongside the model and residuals.\par
The depth from which emission arises varies with emission angle as well as with surface temperature through the temperature dependence of the thermal and electrical skin depth; the ratio of these two skin depths, $\delta_{elec}/\delta_{therm}$, which controls the observed brightness temperature for a given temperature profile, thus varies across Ganymede's surface. A model in which the emissivity varies with emission angle due to refraction at the boundary between the surface and free space according to the Fresnel equations for a smooth dielectric sphere is tested and ruled out because it underpredicts emission at the limb relative to the data. This is consistent with the presence of roughness or volume scattering at millimeter wavelengths; both effects would increase emission detected from high emission angles relative to the smooth model. Note that limb darkening and brightening effects are still present, the former due to the finite thermal inertia (limbs are cooler due to lower insolation) and the latter due to geometric effects (shallower penetration depth for the same path length), but an additional emission-angle dependence in the emissivity is not required. \par
Our model does not currently include volume scattering in the subsurface. For the shallow depths sensed by millimeter observations ($\lesssim$10's of cm) this simplification is appropriate, but interpretation of lower frequency data would likely require a full numerical solution of the radiative transfer equation including the scattering term.
\subsection{Albedo map}\label{sec:albedomap}
The albedo map that was used in the thermal model was produced as follows from \textit{Voyager} and \textit{Galileo} results. Johnson et al. (1983) computed normal albedos at 0.35, 0.41, 0.48, and 0.59 $\mu$m at 24 positions distributed across Ganymede's surface, with corresponding terrain types identified, from \textit{Voyager} observations. We used these albedos to produce normal reflectance spectra for these 24 positions, extending the spectra out to 2.5 $\mu$m wavelengths using Ganymede's disk-integrated spectrum (Clark \& McCord 1980) scaled to match the regions of overlap with the normal albedos. The reflectance spectrum $A_{\lambda}$ is then converted to a wavelength-integrated albedo $A(i)$ via the following ratio
\begin{equation}
A(i) = \frac{\int_{0}^{\infty} A_{\lambda}(\lambda,i)F_{\odot}(\lambda)d\lambda}{\int_{0}^{\infty} F_{\odot}(\lambda)d\lambda}
\end{equation}
where $F_{\odot}$ is the solar spectrum. The 0.35-2.5 $\mu$m window used in this analysis contains over 96\% of the incident sunlight, and we therefore assume that calculating the bolometric albedo from this finite wavelength range does not introduce significant uncertainty. \par
The albedos at these 24 points are compared with the pixel value at their positions on the high-resolution USGS global mosaic\footnote{https://astrogeology.usgs.gov/search/map/Ganymede/Voyager-Galileo/Ganymede\_Voyager\_GalileoSSI\_global\_mosaic\_1km} produced from the best available \textit{Voyager} and \textit{Galileo} images. A polynomial was fit to the albedo as a function of USGS mosaic pixel value, and this polynomial was used to calculate a wavelength-integrated albedo for each pixel in the global mosaic. The scatter of albedos about the best-fit polynomial was 0.045 (1-$\sigma$) and did not show systematic deviations for particular terrain type. Finally, the wavelength-integrated normal albedo map is converted to a bolometric albedo map by multiplying by the phase integral of 0.78 (Buratti 1991), which provides an approximate correction for the shape of the phase function. This approximation assumes that the total absorbed solar power is independent of the incident angle of the sunlight and moreover equates the normal albedo with the geometric albedo, which, like the phase integral, is defined as a disk-integrated quantity (see e.g. Squyres \& Veverka 1982; Young 2017). The correct conversion from normal albedo to absorbed solar power requires knowledge of Ganymede's particle phase function, which is poorly constrained, and for the viewing geometries considered here (solar phase angles $\lesssim 10^{\circ}$ and incident angles $\lesssim 75^{\circ}$) we consider this simplification appropriate. The albedo map is discussed further and shown in Section \ref{sec:results}.
\subsection{Retrievals} \label{sec:retrievals}
Retrieval of the spectral emissivity and near-surface porosity is performed with a simple chi-squared minimization routine using the Nelder-Mead simplex algorithm (Nelder and Mead 1965). The chi-squared value is calculated from the calibrated ALMA image and the corresponding model image after projection of the model onto a sphere and convolution with the ALMA beam. The Nelder-Mead algorithm tends to find local minima, and we initialize the algorithm at a range of starting positions to obtain the global minimum; the global minimum found in this way compares very well with the maximum likelihood parameter values found via Markov chain Monte Carlo (MCMC) simulations, described below. \par
The parameter uncertainties and joint probability distributions are determined from MCMC simulations using the \textbf{emcee} python implementation (Goodman \& Weare 2010; Foreman-Mackey et al. 2013). In the simulations, a set of chains is initialized where each chain is assigned a set of parameter values chosen from a prior distribution. At each timestep, new parameter values are chosen based on the likelihood of the corresponding model as well as the locations in parameter space of the other chains. The multi-dimensional posterior distribution approaches the probability distributions for the parameters given the data, and the uncertainties can be read directly from the posterior distributions. \par
For the work presented here, we initialize 100-200 chains with starting parameters chosen from a gaussian distribution centered on a guess value for each parameter. The integrated autocorrelation time $\tau$ is computed; $\tau$ yields the amount of time (number of steps) that it takes for the chains to forget their previous position, providing both the number of steps needed for the chains to forget their starting position (the burn-in time), as well as the factor by which the number of samples should be reduced to obtain the effective number of independent samples. The autocorrelation time is comparable for all free parameters, and we generate chains that are a minimum of 50$\tau$ in length. The posterior distributions are generated from the remainder of samples in the chains after burn-in. \par
The confidence interval is calculated from the posterior distribution for each parameter by determining the parameter value corresponding to the maximum likelihood estimate, and calculating the interval such that the probabilities of a parameter value falling within the interval above or below the maximum likelihood value are equal (Andrae 2010). That is, for a given parameter $\theta$:
\begin{equation}
\int^{\theta_-}_{-\infty} \textrm{prob}(\theta) d\theta = \int_{\theta_+}^{\infty} \textrm{prob}(\theta) d\theta = (1-0.683)/2
\end{equation}
where 0.683 is chosen for commensurability with the standard deviation.\par
In order to speed up the thermal model computation to make it feasible for the $\sim10^6$ calculations of the model required for each simulation, we employ a surrogate model (sometimes referred to as a response surface model or emulator), or a model of the response as a function of the input parameters that matches the thermal model as closely as possible but with shortened computation time; this approach is common in engineering and other fields (e.g. Queipo et al. 2005). In this case the surrogate model is a multi-dimensional piecewise polynomial constructed from a grid of forward models; the greatest deviation between the thermal model and the surrogate model occurs at high latitudes and at sunrise and sunset, but at all times, locations, and input parameters the accuracy of the surrogate model is better than 0.25 K.\par
\section{Results \& Discussion} \label{sec:results}
We present global best-fit thermal surface properties for Ganymede at frequencies of 97.5, 233, and 343.5 GHz (wavelengths of 3, 1.3, and 0.87 mm), and maps of the temperature residuals from the best-fit global models that indicate systematic trends in thermal properties as well as localized anomalies. The results of the global fits are presented and discussed in Section \ref{sec:res_global}, and the thermal maps presented in Section \ref{sec:res_local}. Section  \ref{sec:ppr} presents a previously-unpublished nighttime infrared observation from the \textit{Galileo} PPR instrument to add context to the discussion. Sections \ref{sec:anomalies} and \ref{sec:res_trends} discuss specific localized anomalies and large-scale trends observed in the collective thermal dataset.
\subsection{Global properties}\label{sec:res_global}
\subsubsection{Brightness temperature and emissivity}
\begin{figure}[ht]
\centering
\includegraphics[width=16cm]{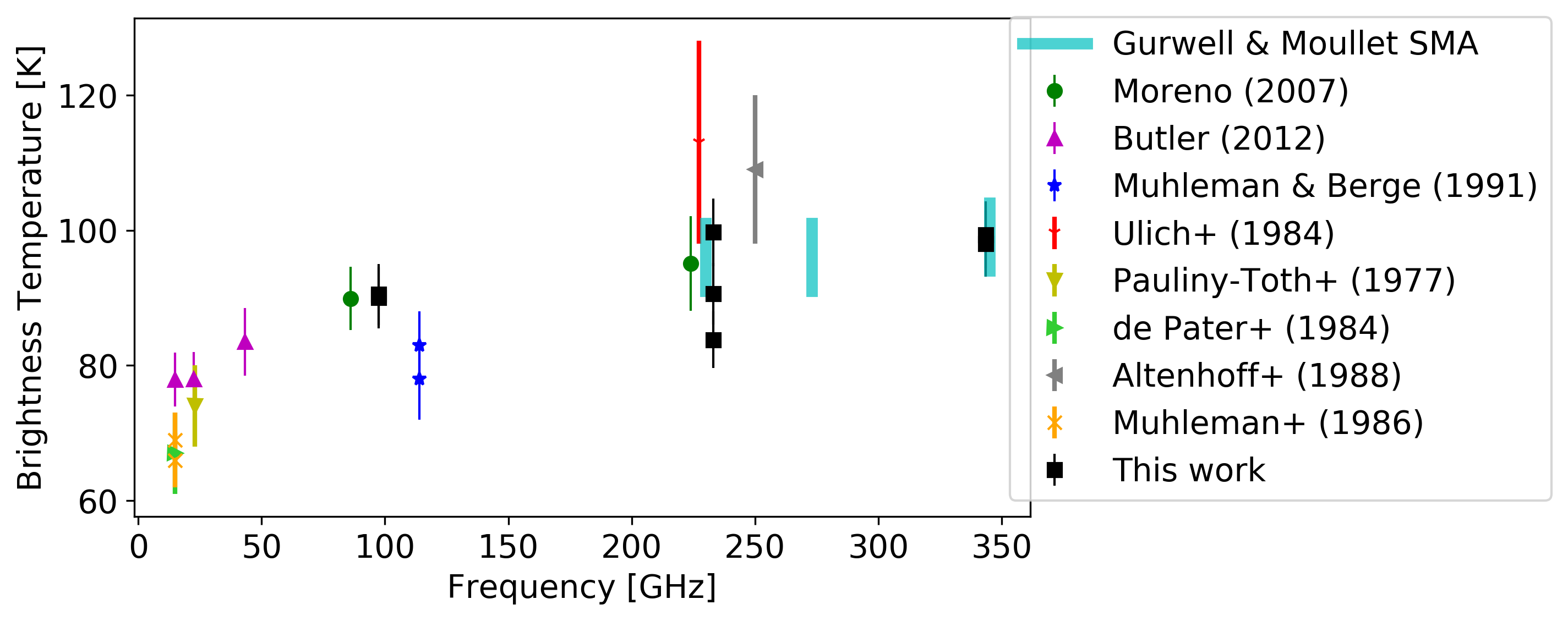}
\caption{Disk-averaged brightness temperature of Ganymede as a function of frequency, from this and past work at cm and mm wavelengths. Past observations are shown from Gurwell \& Moullet (personal communication) SMA observations, where the vertical spread represents the longitudinal variability, as well as from Moreno (2007); Butler (2012); Muhleman and Berge (1991); Ulich et al. (1984); Pauliny-Toth et al. (1977); de Pater et al. (1984); Altenhoff et al. (1988); and Muhleman et al. (1986). \label{fig:Tb_vs_Freq}}
\end{figure}

We find a disk-integrated brightness temperature for Ganymede of 99$\pm$5 K at 343.5 GHz (0.87 mm); of 95$\pm$5 K at 233 GHz (1.3 mm); and of 90$\pm$4 K at 97.5 GHz (3 mm), in good agreement with the majority of past observations in the mm and cm domains (Figure \ref{fig:Tb_vs_Freq}). The corresponding spectral emissivities are 0.78$\pm$0.04, 0.775$\pm$0.04 and 0.75$\pm$0.04 at 343.5, 233, and 97.5 GHz. The uncertainties incorporate flux density scale calibration uncertainties, hemispheric differences, and parameter uncertainties from the modeling. Emissivity is defined here as the ratio of the observed thermal emission to the modeled blackbody emission integrated over the viewing path (described in Section \ref{sec:model}), rather than emission from just the surface. If the latter definition were used, the emissivities would still fall within the 0.7-0.8 range but would show a stronger frequency dependence because the electrical skin depth is larger at the lower frequencies. \par
While the frequency dependence of Ganymede's disk-integrated brightness temperature matches well with the trend seen in previous observations, Muhleman \& Berge (1991) found an anomalously low brightness temperature for Ganymede at 115 GHz, near in frequency to our Band 3 observations. Figure \ref{fig:Tb_vs_Freq}, which places our measurements in the context of disk-integrated measurements from 10 through 350 GHz over the past several decades, shows that the measurements of Muhleman and Berge (1991) are inconsistent with all neighboring measurements including ours. Excepting their observation, the measurements collectively indicate a brightness temperature that is slightly higher at 350 GHz than at 230 GHz, but only begins to drop off in earnest below 100 GHz and decreases roughly linearly in frequency below 100 GHz from $\sim$90 K to $<$80 K by 10 GHz. Over the frequencies covered by our ALMA dataset, the decrease in brightness temperature can be accounted for by the fact that lower frequencies are sensitive to emission from deeper in the surface, through the compounding effects of colder physical temperatures at depth, and of the higher thermal inertias of deeper layers resulting in lower (daytime) brightness temperatures. \par
The millimeter emissivities found for Ganymede are consistent with emissivities measured at these frequencies for dry compacted snow and ice on Earth near freezing temperatures, although the absolute emissivity ranges from 0.6-1.0 in such studies and its frequency dependency may reverse depending on the exact characteristics of the snow/ice (e.g. Hewison and English 1999; Yan et al. 2008). While our best-fit emissivities are slightly lower at lower frequencies, this variation is within the uncertainties, although as noted above, if we were to define emissivity relative to surface temperature instead of accounting for subsurface sounding the emissivities would increase with frequency. ALMA observations of the Pluto-Charon system also found lower emissivities at lower frequencies, with values in the range of 0.7-0.9 similar to what we find for Ganymede (Lellouch et al. 2017). Among the icy Galilean satellites, published ALMA data only exist for a single frequency of 233 GHz of Europa; the emissivity of Europa at that frequency matches what we find for Ganymede (Trumbo et al. 2018), although the emissivity for Europa is defined relative the surface rather than subsurface emission, which results in a lower derived value. \par
\subsubsection{Thermal inertia and porosity}
Our best-fit global models have porosities of $35\pm25$\%, $45\pm30$\%, and $10^{+30}_{-10}$\% for data at 343.5, 233 GHz, and 97.5 GHz (0.87, 1 mm, and 3 mm) respectively, corresponding to effective thermal inertias of $\Gamma_{\text{eff}}=450^{+300}_{-250}$, $350^{+350}_{-250}$, and $750^{+200}_{-350}$ J m$^{-2}$ K$^{-1}$ s$^{-1/2}$ for the adopted conductivity model. These values, and all residual maps, are based on joint fits to the available datasets in each frequency band. Individual images alone are unable to provide robust constraints on the thermal properties; MCMC simulations for individual datasets find small mathematical uncertainties, but the best-fit parameter values are susceptible to localized variations in thermal properties across the disk and vary between observations at different viewing geometries. The large uncertainties on the global best-fit values, which are derived from the joint MCMC simulations, arise from the fact that at these low porosities/high thermal inertias, and in particular at the depths probed by the observations, the diurnal temperature variations are already low so that changes in the parameters produce only modest changes in the diurnal temperature variations. Moreover, a lower porosity surface is less conductive but more transparent to emission, so the effects of changing porosity on the global brightness temperature distribution partly cancel out, particularly at the lowest frequency. \par
Past infrared observations of Ganymede have been fit with a thermal inertia of 70; however, a single thermal inertia has not fit these past data well and a two-component model provided a better fit (Morrison and Cruikshank 1973; Spencer 1987; Pappalardo et al. 2004). In the two-component model, the best-fit thermal inertias were found to be around 20 and 500$\pm$100, but the data were unable to differentiate between models in which the components were vertically vs. horizontally segregated. Our millimeter observations are sensitive to deeper layers in the subsurface than the infrared observations; we performed tests using a two-component model with horizontal segregation, and the fits to the millimeter data were not improved by the addition of a low-$\Gamma$ component. These results therefore support the vertical rather than horizontal stratification scenario. The increase in thermal inertia from 233 and 343.5 GHz ($\Gamma=400$) to 97.5 GHz ($\Gamma=750$) is also consistent with this inference but suggests a thermal inertia gradient with depth rather than a two-layer model with distinct high and low thermal inertia components. Note that solid ice or rock has a thermal inertia even greater than the high thermal inertia component in these models ($\Gamma=2000$ for solid ice). \par
Under the simplifying assumption that the infrared emission arises from the surface and the millimeter emission arises from the upper one electrical skin depth, the collective infrared and millimeter constraints provide constraints on porosity variation with depth. Adopting the functional form for that dependence given in Equation \ref{eqn:rho_vs_z}, the data imply a decrease in porosity from $\sim$85\% at the surface to 10\% at roughly a half meter depth, over a compaction length scale of tens of cm. This qualitative increase in density and thermal conductivity with depth is robust to model choices. However, the exact values for these porosities would change if a different grain size or conductivity model were used. In particular, if the surface ice were amorphous or had looser grain contacts than the deeper ice, a lower surface porosity would be needed and hence a shallower porosity gradient. \par
A higher thermal inertia at longer wavelengths, sensitive to deeper in the near sub-surface, has also been derived from observations of other icy satellites in the Solar System, as well as Io (de Pater et al. 2020). The thermal inertia of Europa near 233 GHz is 95 (Trumbo et al. 2018), which is lower than we find for Ganymede, but this difference is consistent with trends in the infrared data, which also find lower thermal inertias for Europa ($\Gamma=50$ for a 1-component model or 15 and 300$\pm$200 for a 2-component model; Spencer et al. 1987). Thus for both Europa and Ganymede, the millimeter thermal inertia is consistent with the high thermal inertia component of the 2-component infrared models, supporting the hypothesis of a vertical compaction gradient or stratification in the surface, with a high thermal inertia component underlying a low thermal inertia veneer of at most a few mm thickness. The same effect has also been seen on the saturnian satellites from a combination of infrared and centimeter-wavelength \textit{Cassini} observations (e.g. Le Gall et al. 2014; Bonnefoy et al. 2020), although the thermal inertia is higher on the jovian than the saturnian satellites for a given frequency band. \par
The depth to which observations at millimeter wavelengths are sensitive in an icy surface is a strong function of temperature, frequency, and composition; pure, porous water ice is extremely transparent. Even for the higher ALMA frequencies (at or above 230 GHz; wavelength below 1 mm), emission may arise from below a diurnal skin depth, where the temperature may differ substantially from the surface temperature to a degree that varies between terrains. This effect is included in our model using the measurements available for ice and dusty dry snow, but the general lack of appropriate experimental data on the dependence of dielectric properties on density and impurities at the relevant temperatures and frequencies is a source of uncertainty. In addition, scattering within the subsurface become increasingly relevant for emission arising from a greater depth; inclusion of these processes will be relevant for lower frequency data. \par
These modeling simplifications, as well as choices for parameterizations of the dependence of thermal properties on emission angle, density, and temperature, has an effect on the best-fit global properties. During model development, several different model parameterization choices were tested, and the spread of values give some sense of the uncertainty on properties resulting from model assumptions. Over a range of modeling choices, the best-fit thermal inertias still fall within or close to the range covered by our uncertainties. Above a thermal inertia of a few hundred the predicted temperature distribution is only weakly sensitive to changes in thermal inertia, and much higher thermal inertias cannot be completely ruled out. 
\subsection{Thermal maps}\label{sec:res_local}
\begin{figure}[ht]
\centering
\includegraphics[width=10cm]{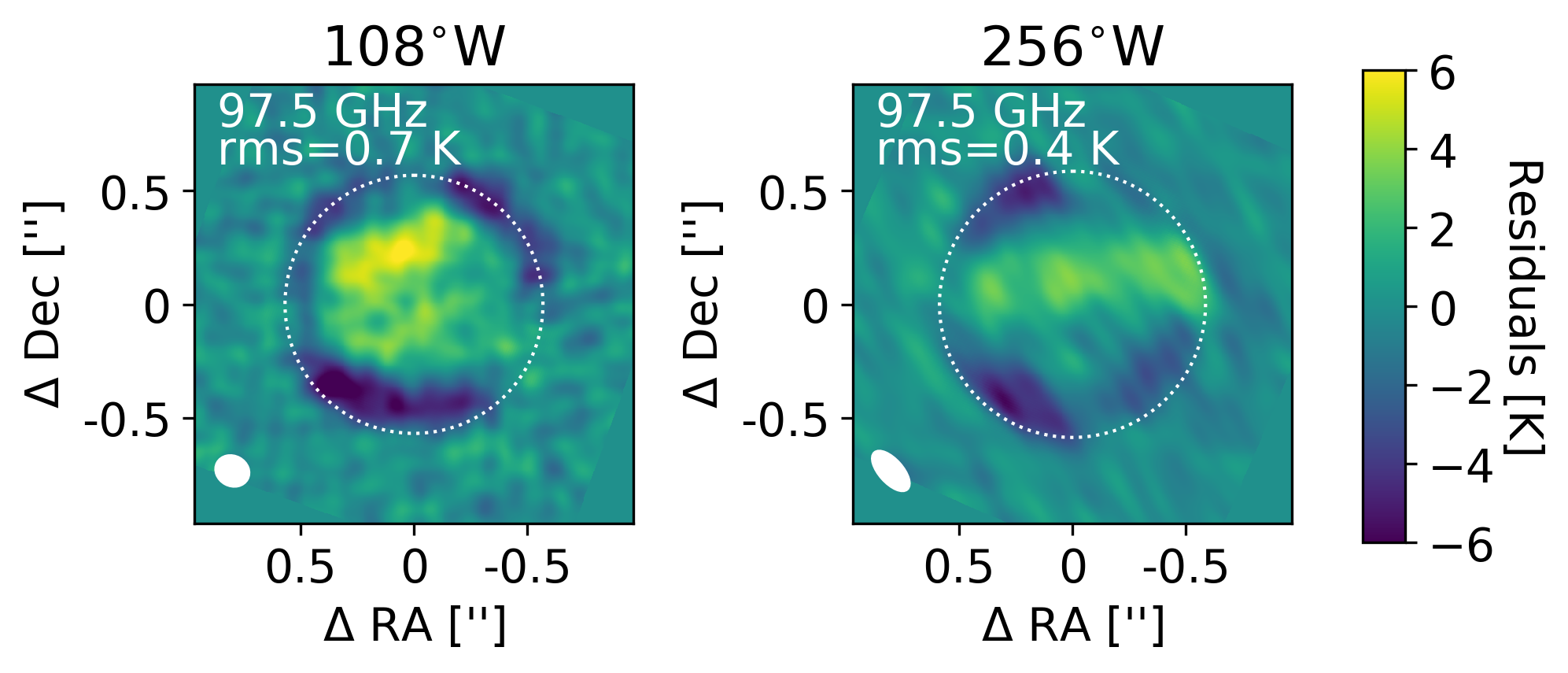}
\includegraphics[width=14cm]{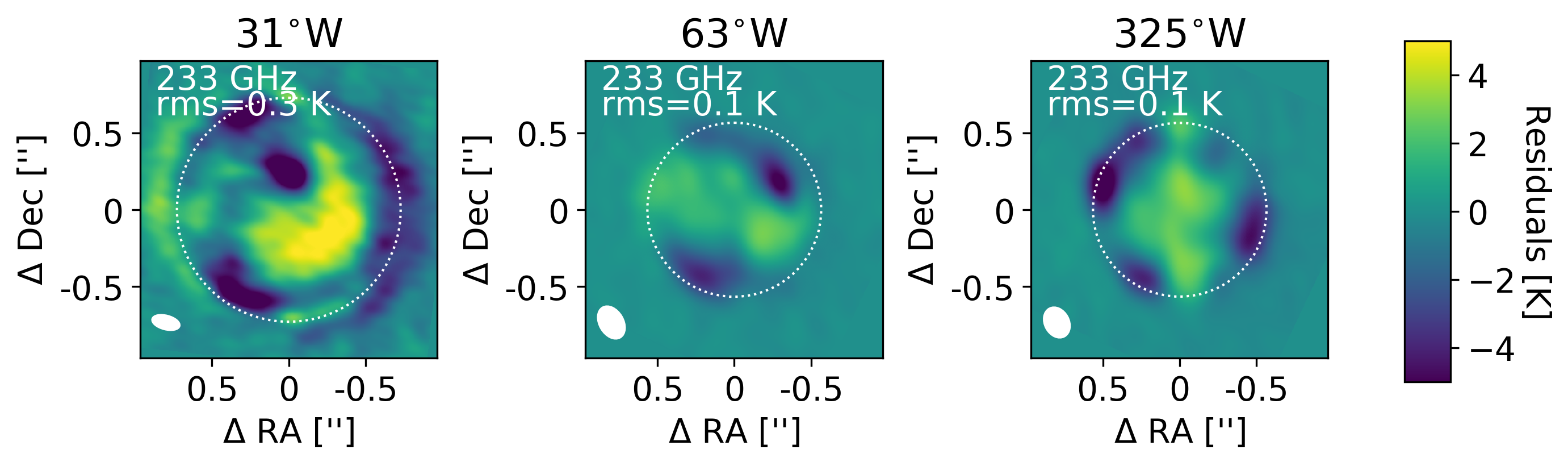}
\includegraphics[width=10cm]{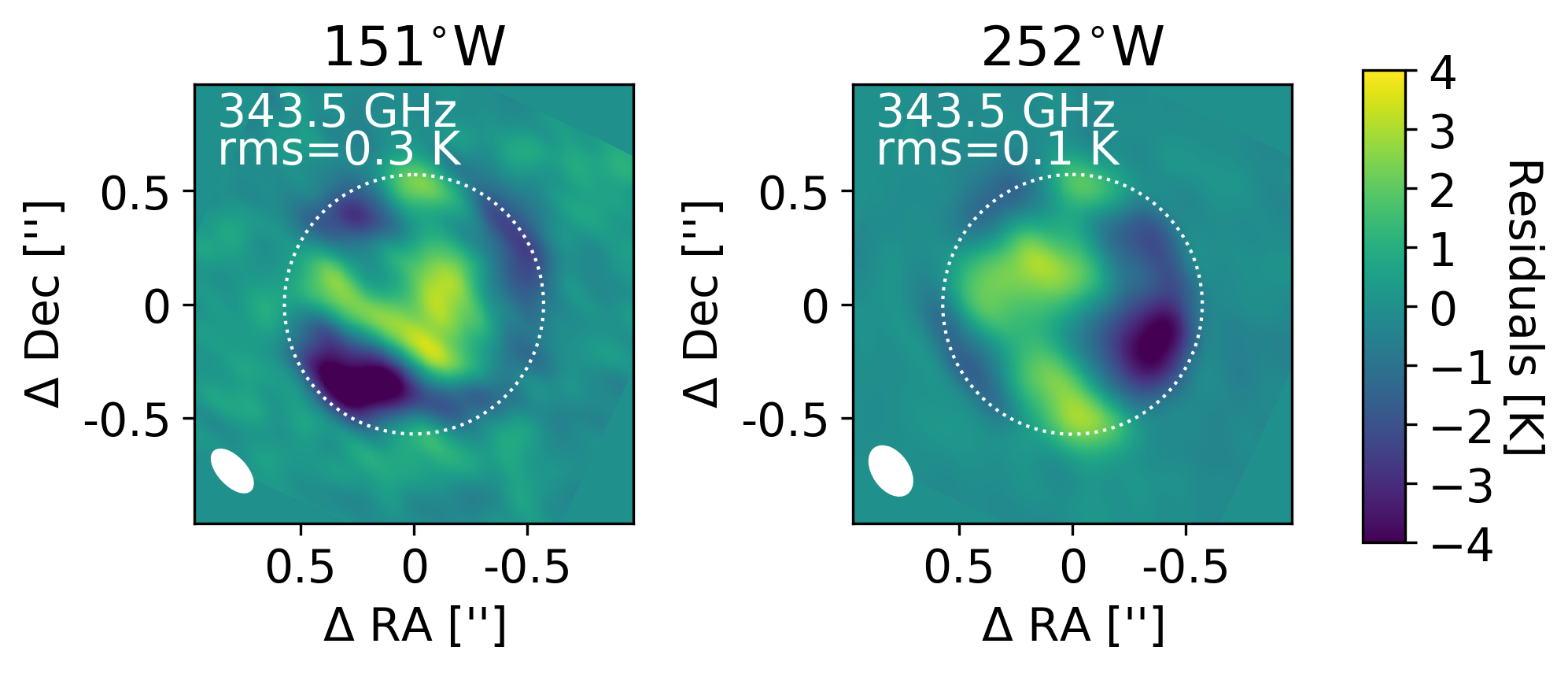}
\caption{Residuals (data-model) using a model with porosities of 10\%, 43\%, and 34\%, the best-fit value for each frequency band, corresponding to effective thermal inertia of $\Gamma_{\text{eff}}=$750, 350, and 450, at frequencies of 97.5, 233 and 343.5 GHz respectively. The structured pattern in the images is an artifact arising from performing a deconvolution with incomplete uv coverage. \label{fig:residuals}}
\end{figure}

\begin{figure}[ht]
\centering
\includegraphics[width=16cm]{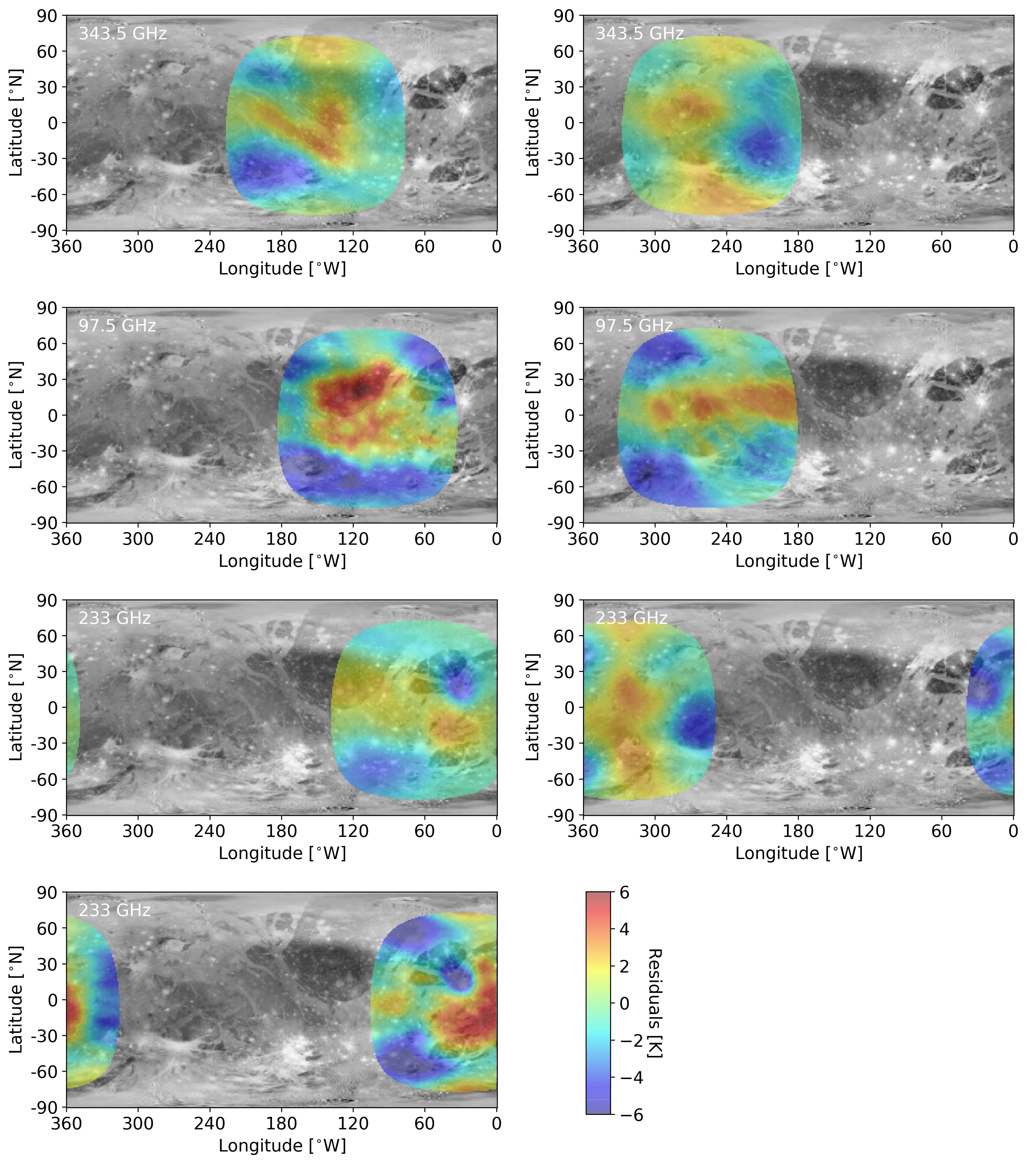}
\caption{Residuals (data-model) after the subtraction of best-fit models as shown in Figure \ref{fig:residuals}, shown in units of brightness temperature and in a cylindrical projection in overlay on a Ganymede albedo map, cropped to include only regions with an emission angle under 75$^{\circ}$. The scaling is the same on all images and corresponds to the colorbar. Ganymede's leading and trailing hemispheres span longitudes 0-180$^{\circ}$W and 180-360$^{\circ}$W respectively. \label{fig:Tmaps}}
\end{figure}

Localized residuals from the global best-fit models are present in all datasets well above the noise level ($>5\sigma$ if $\sigma$ is derived from the off-disk sky background). Figure \ref{fig:residuals} shows the residuals from each observation after subtracting the best-fit model for that frequency. The residuals are shown in cylindrical projection in overlay on the albedo map of Ganymede in Figure \ref{fig:Tmaps}. Despite the caveats noted above for the global best-fit parameters, the local anomalies are fairly robust to changes in the global properties and modeling assumptions. In addition, the regions that are hotter or colder than the global best fit are typically consistent between observations made at different viewing geometries, supporting the interpretation of these features as regions of different thermal properties rather than data artifacts.\par
The projected residual maps from all seven observations are combined, averaging together in areas of overlap, in Figure \ref{fig:ResMap}. By combining observations we cover all longitudes on Ganymede's surface, although we note that the observations are made at three different frequencies sensitive to somewhat different depths, and moreover combine data taken of a given region at different times of day, and the map should therefore be viewed with some caution. These temperature residuals are shown alongside the albedo map in Figure \ref{fig:AlbedoComp}, where it can be seen that although some albedo features have corresponding temperature residual features, the correspondence is limited to certain terrains. In addition, the degree of temperature residuals are too great to be explained by modest changes in albedo: a 10\% error in albedo could only account for less than half of the temperature residual seen at Tros, and a quarter of the residual in Galileo Regio. We therefore attribute the residuals to localized physical differences rather than to systematic errors in the assumed albedos. \par
\begin{figure}[ht]
\centering
\includegraphics[width=16cm]{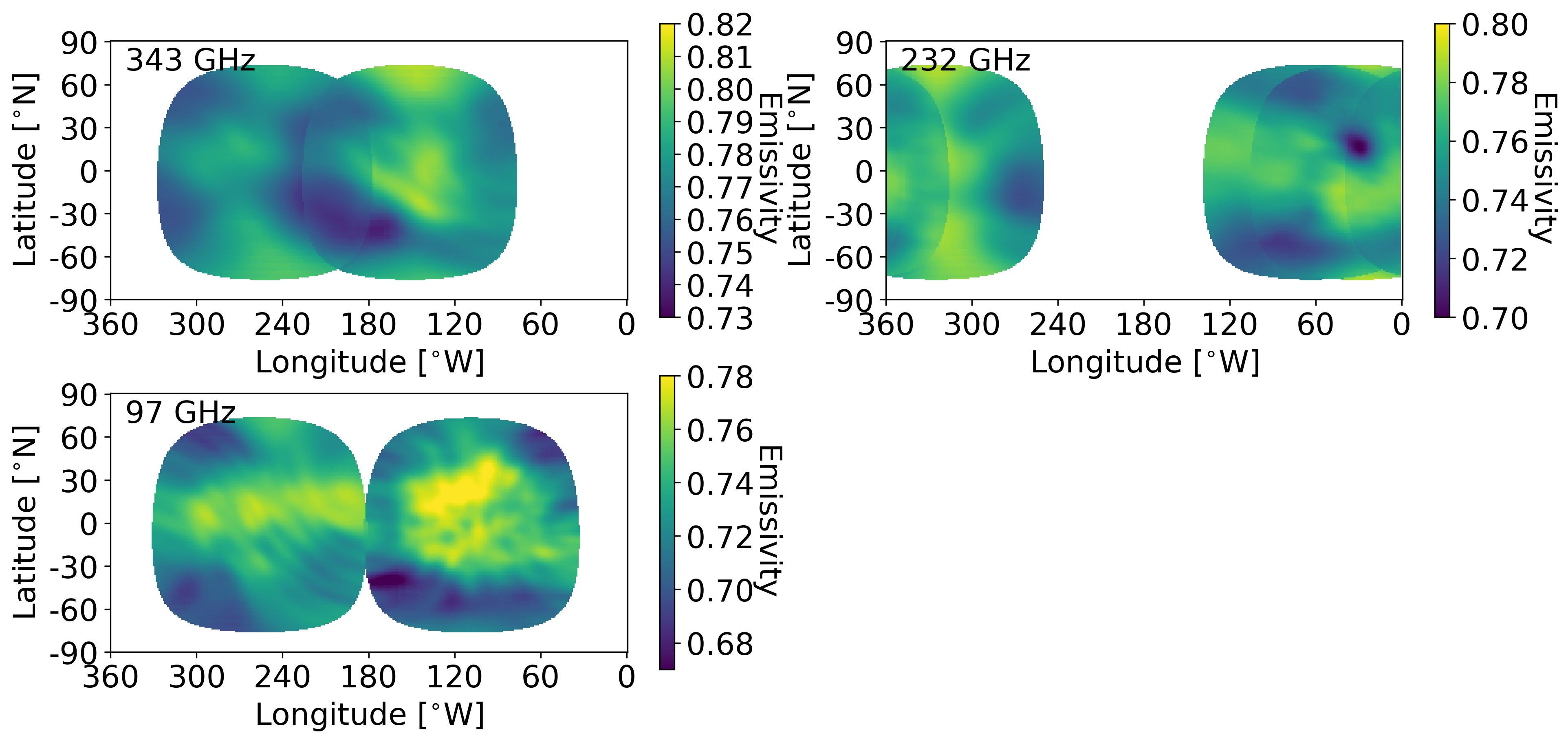}
\caption{Maps of emissivity across Ganymede's surface at the three frequencies of observation, under the scenario in which all temperature residuals from the global best fit models arises from spatial variations in emissivity alone. Porosities of 10\%, 43\%, and 34\% are adopted at frequencies of 97.5, 232 and 343.5 GHz respectively, corresponding to effective thermal inertias of $\Gamma_{\text{eff}}=$750, 350, and 450. The average emissivity of each 232 GHz observation is adjusted to match the average across all three observations to obtain a match in overlapping regions. \label{fig:Emaps}}
\end{figure}

 
The temperature residuals arise from spatial variations in thermophysical properties and/or millimeter emissivity, both of which can vary with composition, porosity, and grain properties. Ganymede's terrain is heterogeneous at a scale much smaller than the spatial resolution of the data, and higher resolution observations than are possible from the ground, as well as nighttime observations, would be required to robustly separate the effects of different thermal properties. Figure \ref{fig:Emaps} demonstrates the quantitative variations in emissivity that would be required to fit the residuals if surface porosity were assumed to be fixed globally. A converse analysis, in which emissivity is fixed at the global best-fit value and surface porosity varied locally, is unable to match the full range of temperature residuals. In particular, localized cold regions are colder than predicted for any parameter values if the emissivity is fixed, while certain dark terrains and the equatorial limbs are too warm in some datasets.
\subsection{Comparison with Galileo PPR nighttime observation}\label{sec:ppr}
The Photopolarimeter-Radiometer (PPR) instrument (Russell et al. 1992) on the \textit{Galileo} spacecraft obtained a large set of observations of Ganymede's surface during the day as well as at night, collectively covering much of the surface at a range of spatial resolutions. A full analysis of that dataset is not the focus of this work, but we analyze a single, previously-unpublished thermal observation from the G7 orbit using the $>$42 $\mu$m filter. This observation is unique in that it has the best global nighttime coverage of Ganymede, covering longitudes from 120-240$^{\circ}$W and latitudes nearly from pole to pole. This observation complements the ALMA dataset in both wavelength and time-of-day coverage, and we include it to aid in interpretation of the ALMA observations. \par
The PPR data have been reduced through the standard pipeline, de-boomed (corrected for obscuration by the spacecraft boom), and corrected for non-zero sky levels. Figure \ref{fig:PPR}a shows the nighttime temperature map from PPR. As has been found by previous authors for infrared data (see Section \ref{sec:intro}), we find that a 2-component thermal model with horizontal-segregated components provides a better fit than a 1-component model, with the best fits being obtained for thermal inertias of $\Gamma=50$ and $\Gamma=400-600$. Such a model is shown in \ref{fig:PPR}b, and the residuals from the model in Figure \ref{fig:PPR}c. We note that in modeling infrared data we neglect subsurface emission and fit for a thermal inertia value that is fixed across the surface; under these conditions, our model is very similar to past Galilean satellite models (e.g. Spencer et al. 1989), and the match between our results and past results by other modelers is a useful validation. \par
The infrared data show a clear correlation between thermal properties and terrain type: the dark terrain is warmer than the grooved terrain during the day but cooler than the grooved terrain at night, indicating that the grooved terrain has a higher thermal inertia (Pappalardo et al. 2004). This is also clearly seen in the model residuals in Figure \ref{fig:PPR}c: the grooved terrain is warmer than expected at night after accounting for albedo, for a model that assumes all terrains have the same thermal properties. The impact crater Osiris is clearly colder than the surrounding grooved terrain at night, but is still slightly warm relative to the best-fit model, suggestive of a thermal inertia intermediate between the dark and bright terrain, perhaps with different emissivity properties (Figure \ref{fig:PPR}). \par
A comparison between the PPR nighttime and ALMA daytime temperature residuals show some suggestion of the expected day/night anti-correlation in the equatorial regions, between $\pm$30$^{\circ}$ latitude. That is, regions with higher thermal inertia are cooler during the day and warmer at night than the disk average, while regions with lower thermal inertia are warmer during the day and cooler at night than the average. This effect can be seen most convincingly in Galileo Regio (labeled on Figure \ref{fig:AlbedoComp} for reference) and the dark terrain southwest of it, as well as the grooved channel between the two. In these regions, the datasets together are broadly consistent with the thermal inertia terrain correlations described above, which would imply that material properties measured in the infrared extend down to at least the tens of cm depths that ALMA is sensitive to. \par
However, this conclusion is not supported in other terrains, where the ALMA temperature residuals show little resemblance to the PPR data. In addition, while the ALMA data show some correlation between positive temperature anomalies and dark terrains, this correlation is not as strong or consistent as in the PPR data. A good example of this is the region of dark terrain near 210$^{\circ}$W and 0-30$^{\circ}$S, which is one of the most disrupted areas of dark terrain and appears anomalously cold in the ALMA data despite behaving as typical dark terrain in the PPR data.
\subsection{Localized thermal anomalies}\label{sec:anomalies}
While the overall correlation between thermal anomalies and dark vs. grooved terrain is weak, there are several specific surface terrains that do appear correlated with positive and negative temperature anomalies. The dark terrains in the leading hemisphere are generally associated with positive anomalies, while the bright impact craters are associated with negative anomalies. A few particular examples will be discussed in more detail below.
\subsubsection{Galileo Regio} \label{sec:res_galregio}
\begin{figure}[ht]
\centering
\includegraphics[width=8cm]{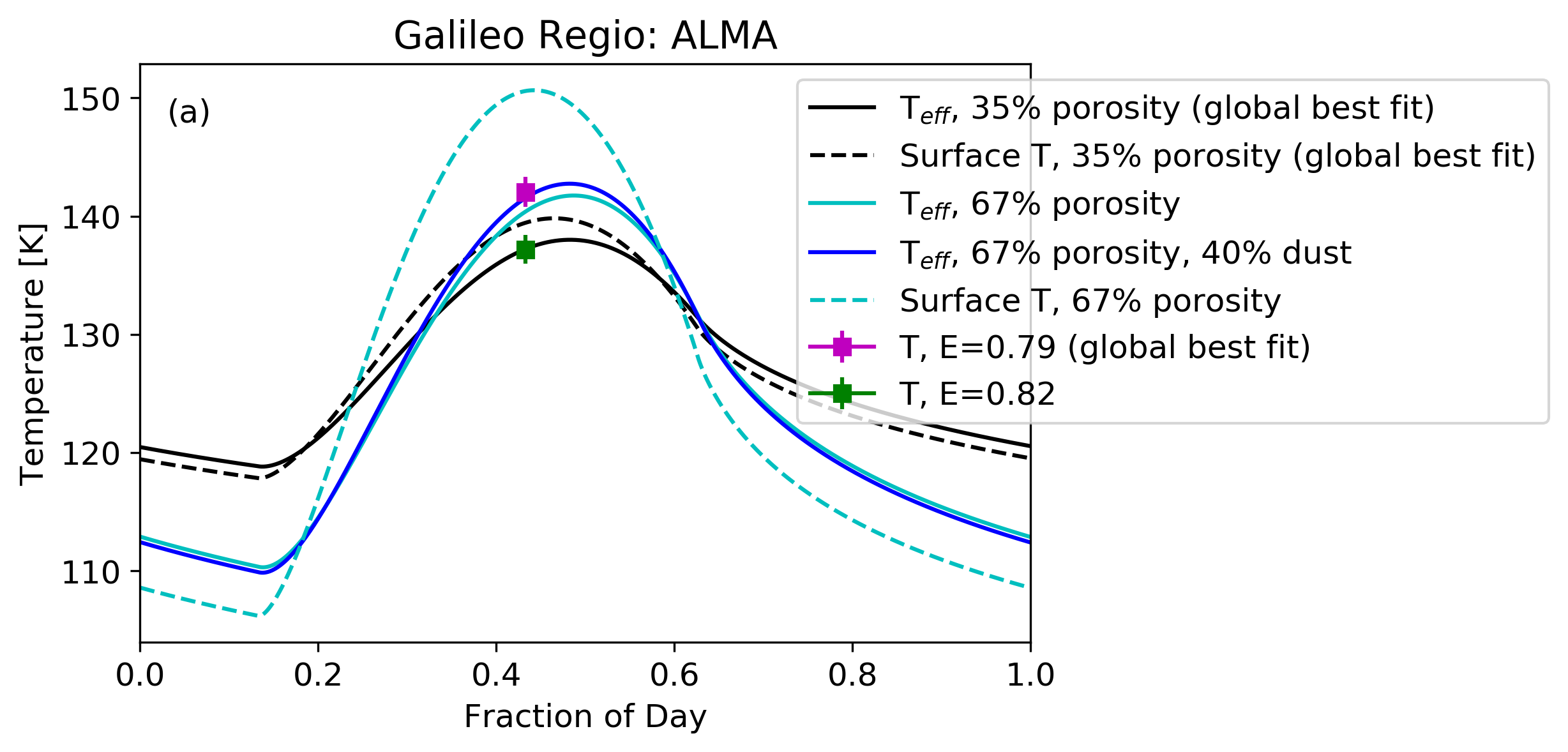}
\includegraphics[width=8cm]{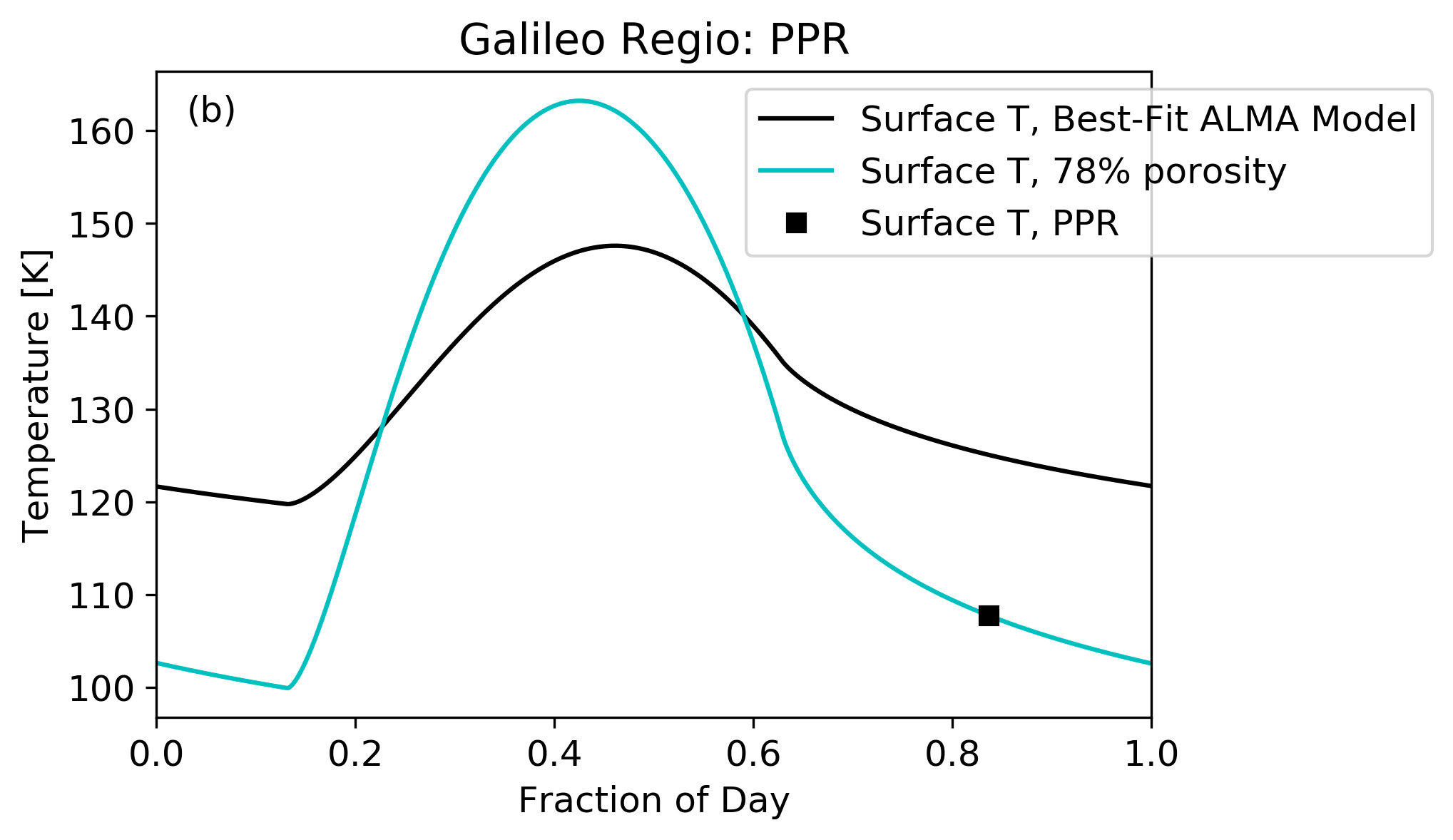}
\includegraphics[width=8cm]{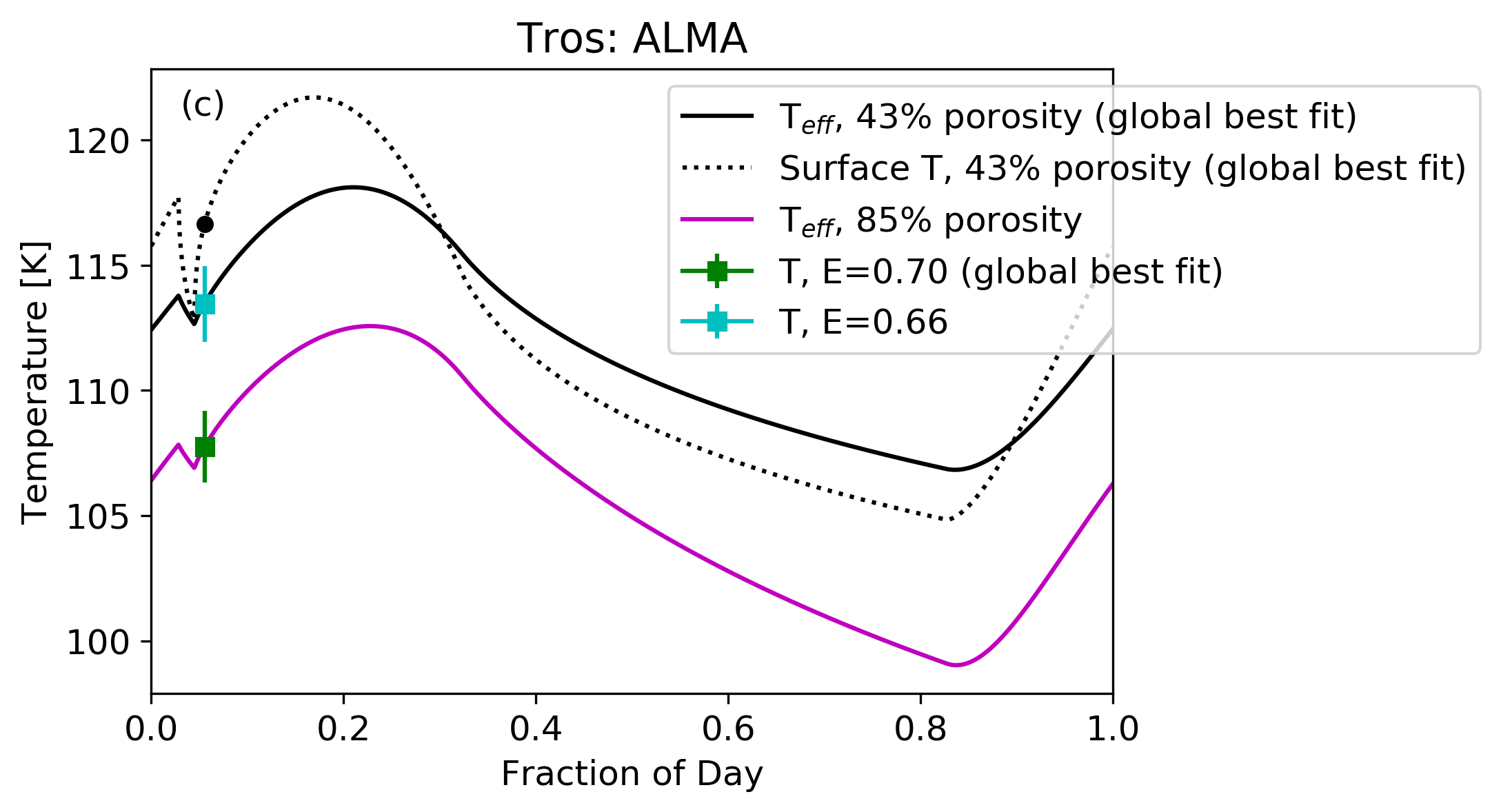}
\caption{Temperature as a function of time of day for representative low albedo (Galileo Regio) and high albedo (Tros) regions on Ganymede's surface that show temperature anomalies: (a) Models of the temperature of Galileo Regio using the global best-fit model (black), a model that has a localized higher porosity in this region (cyan) and a higher porosity plus higher dust fraction (blue). The effect of changing emissivity on the measured temperature is also shown. Note that the ALMA data should be compared with the $T_{\text{eff}}$ curves; surface temperature is shown for context. (b) Models of the surface temperature in Galileo Regio compared to \textit{Galileo} PPR, for a model in which the surface thermal properties are the same as those at the depth ALMA is sensitive to (black) vs. with a higher porosity at the surface (cyan). (c) Similar to (a) but for the high-albedo impact crater Tros. The circles on the temperature curves show the modeled temperature at the time of observation. Although the data appear to fall close to eclipse, they were taken several hours after eclipse end. \label{fig:Tvst}}
\end{figure}

\begin{figure}[ht]
\centering
\includegraphics[width=8cm]{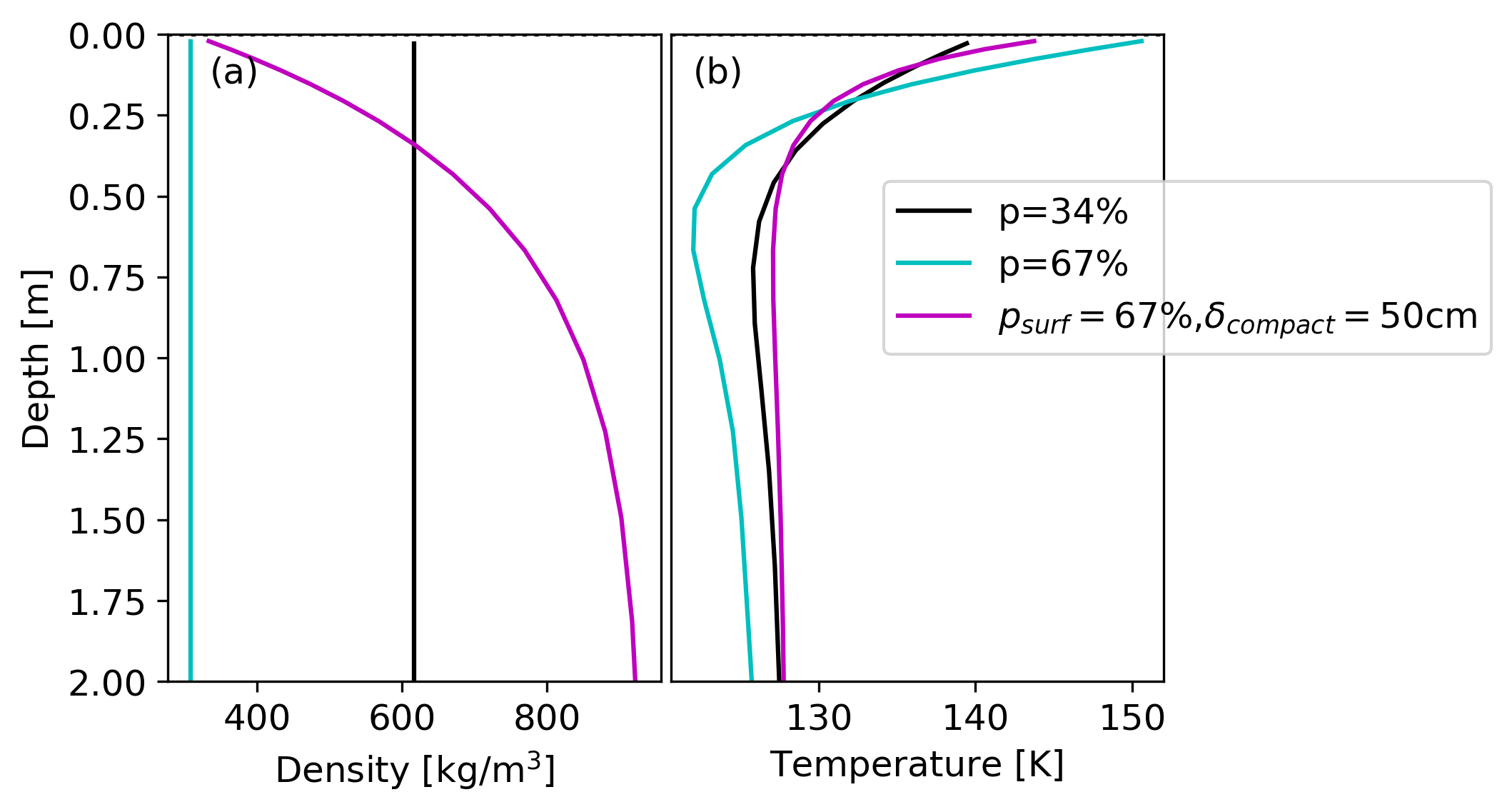}
\includegraphics[width=8cm]{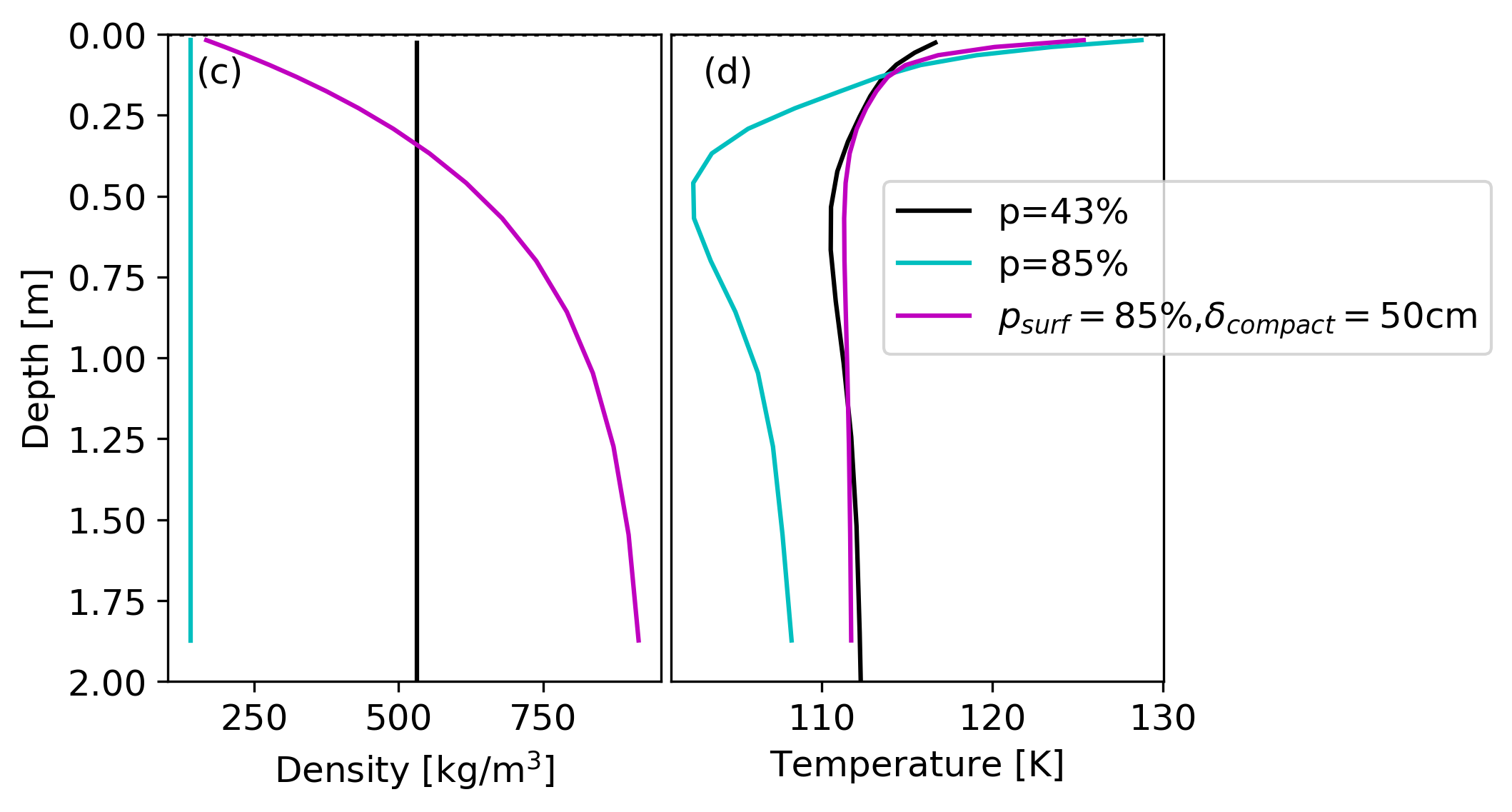}
\caption{Afternoon temperature as a function of depth for anomalous warm and cold regions (b) Galileo Regio and (d) Tros, for the density profiles shown in (a) and (b). Three cases are shown for each: the global best fit porosity for the frequency of observation covering that region in the afternoon, the local variation in porosity that produces the best fit to the data (see Figure \ref{fig:Tvst}), and a case where the porosity as a function of depth is characterized by the surface porosity and compaction length scale given in the legend (see Equation \ref{eqn:rho_vs_z}).\label{fig:Tz}}
\end{figure}

Galileo Regio is a large region of dark terrain in Ganymede's northern hemisphere, spanning from near the equator up to the north pole. In the ALMA data, this region and the other leading-hemisphere dark terrains are the most prominent examples of positive thermal anomalies associated with geological terrains. The southern part of Galileo Regio in particular (the portion of Galileo Regio not covered by the polar caps) shows the greatest positive temperature anomaly in Figure \ref{fig:ResMap}. The degree of the anomaly cannot be explained by an underestimate of albedo (a 10\% underestimate could only account for roughly a quarter of observed temperature anomaly). The PPR and ALMA data together suggest a lower thermal inertia in this region. However, lowering thermal inertia by raising the surface porosity is not sufficient to match the heightened temperature; increasing porosity also raises surface transparency allowing emission from greater depths, and the maximum surface temperature that can be obtained comes from a balance between these two effects. That maximum surface temperature is shown in the temperature curve in Figure \ref{fig:Tvst} that corresponds to an increased porosity alone, and does not quite match the observed temperature. However, raising the porosity and increasing the dust mass fraction (which decreases transparency) can together reproduce the observed temperature. Figure \ref{fig:Tz} shows the vertical temperature profiles corresponding to these two porosity cases, as well as a case with depth-dependent porosity, and gives a sense for the magnitude of subsurface temperature gradient.\par
Although dark terrains are associated with some positive anomalies, the correspondence is much less clear than that seen in the thermal infrared in \textit{Voyager} and \textit{Galileo} data (Pappalardo et al. 2004; Section \ref{sec:ppr}). This might suggest that the millimeter data are ``seeing through'' a thin veneer of material that the infrared data are sensitive to, and that the thermal properties at a depth of a few centimeters are not correlated with terrain type. However, it is unlikely that we are seeing through the dark terrain material itself: a comparison of slope angles and crater morphologies between dark and bright terrains indicates that the lag deposit producing the dark coloration should be on the order of few meters thick (Pappalardo et al. 2004), much deeper than millimeter wavelengths are sensitive to. 
\subsubsection{Bright impact craters} \label{sec:res_craters}
\begin{figure}[ht]
\centering
\includegraphics[width=14cm]{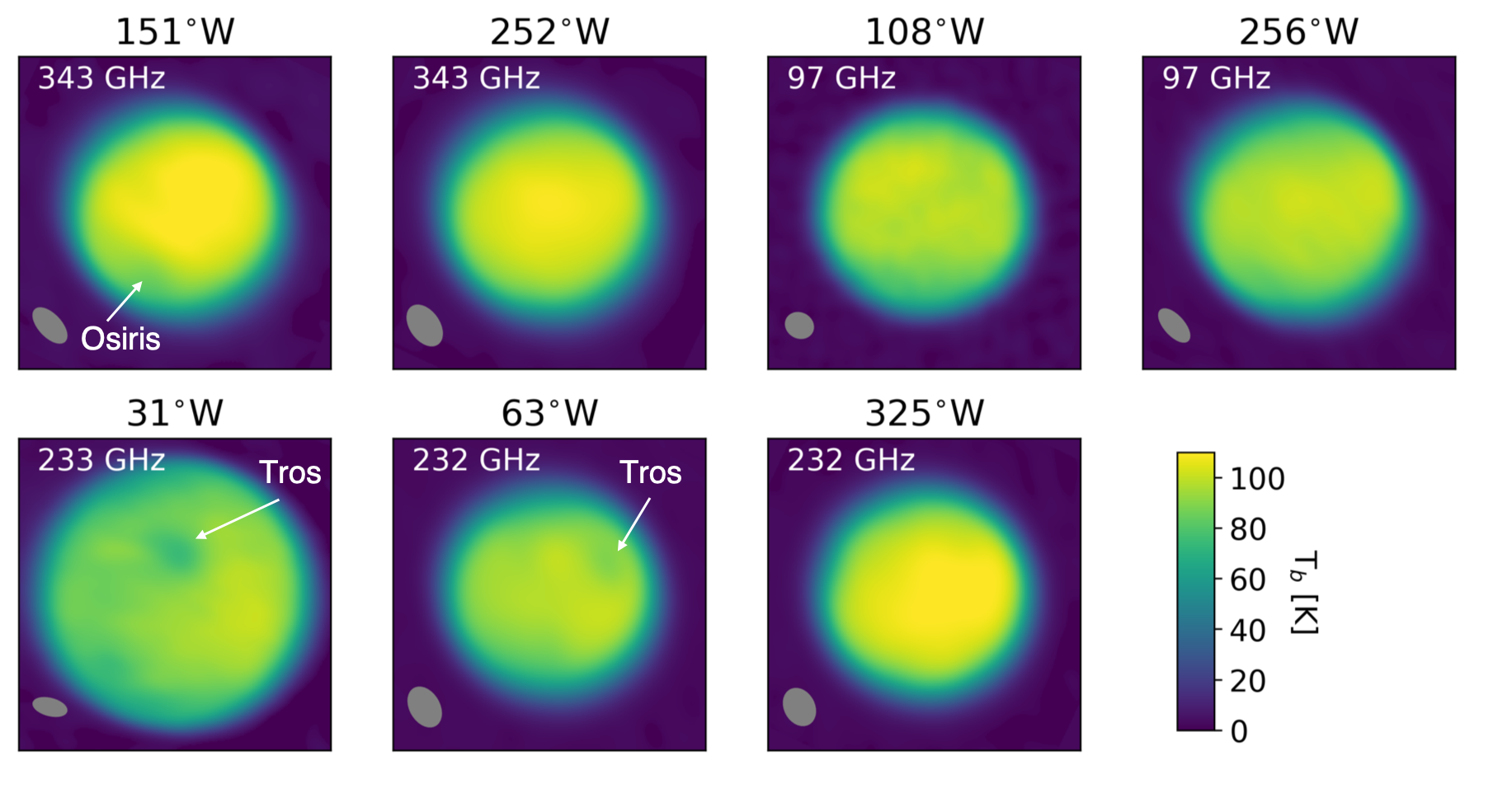}
\caption{Ganymede ALMA observations, with bright (cold) impact craters Tros and Osiris labeled. All data have been calibrated and rotated so that Ganymede's north pole is up. The central meridian longitude and frequency of observation are indicated on each panel, and the gray ellipse represents the major and minor full-width at half maximum and position angle of the restoring beam, which is the effective resolution of the observation. The brightness temperatures and angular scales are the same for all images; the angular size of Ganymede is set by the Earth-Jupiter distance at the time of observation (Table \ref{tbl:obs}).\label{fig:data}}
\end{figure}

\begin{figure}[ht]
\centering
\includegraphics[width=16cm]{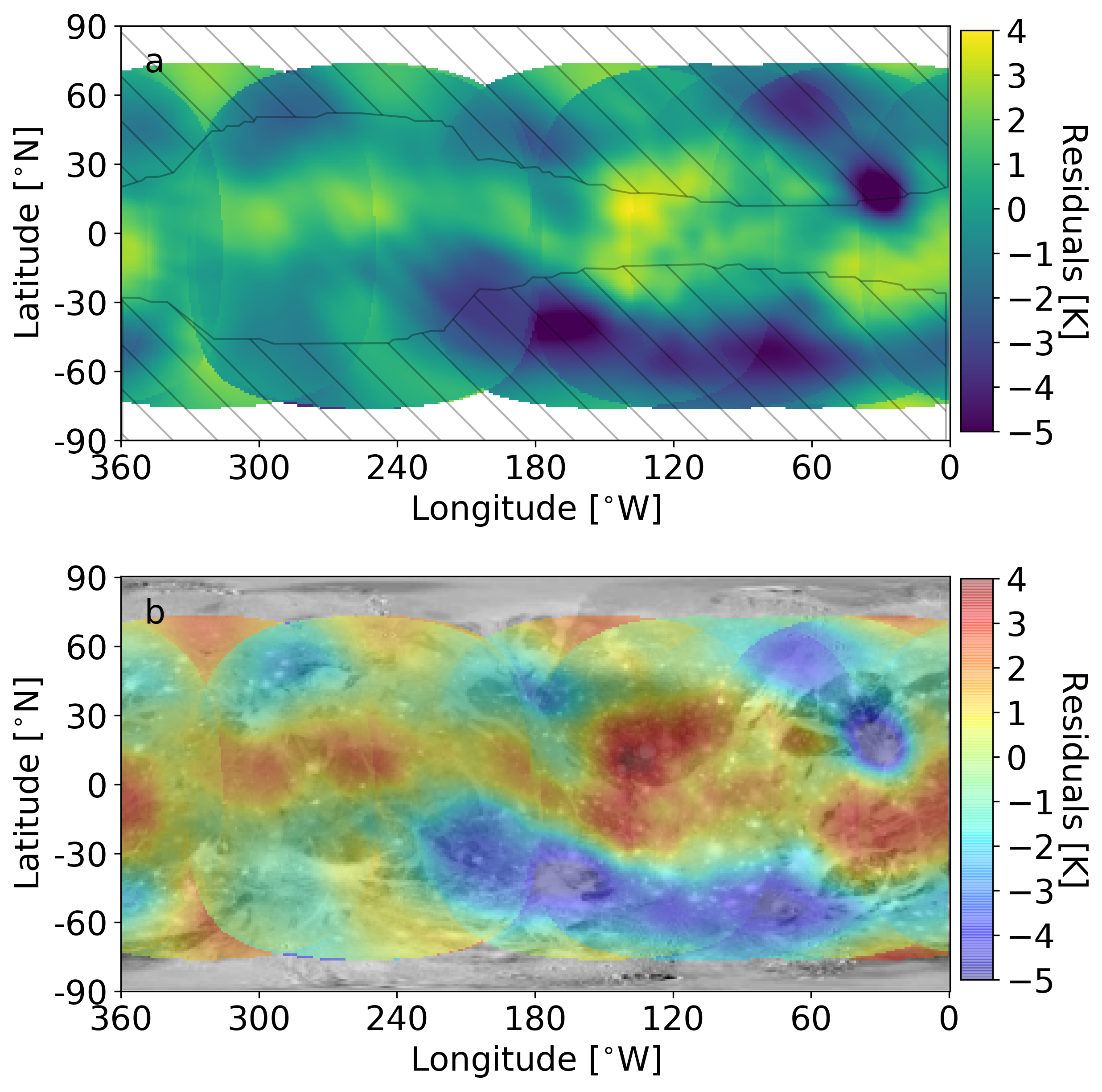}
\caption{Temperature residuals (data-model) after subtraction of best-fit models averaged across all observations. (a) Combined residuals from Figure \ref{fig:Tmaps}, with surface regions with emission angle greater than 75$^{\circ}$ excluded; and (b) the same map in overlay on the albedo map. The hatched regions in (a) indicate the locations of where plasma flux reaches the surface from Jia et al. (2008; 2009), which provides a good match to the brightest regions of Ganymede's UV aurora (McGrath et al. 2013). Ganymede's leading and trailing hemispheres span longitudes 0-180$^{\circ}$W and 180-360$^{\circ}$W respectively. Non-zero temperature residuals indicate localized geographical regions with anomalous thermal properties, although these maps should be viewed with some caution because they combine data from three frequencies sensitive to different depths, and moreover combine the same surface regions observed at different times of day. \label{fig:ResMap}}
\end{figure}

\begin{figure}[ht]
\centering
\includegraphics[width=12cm]{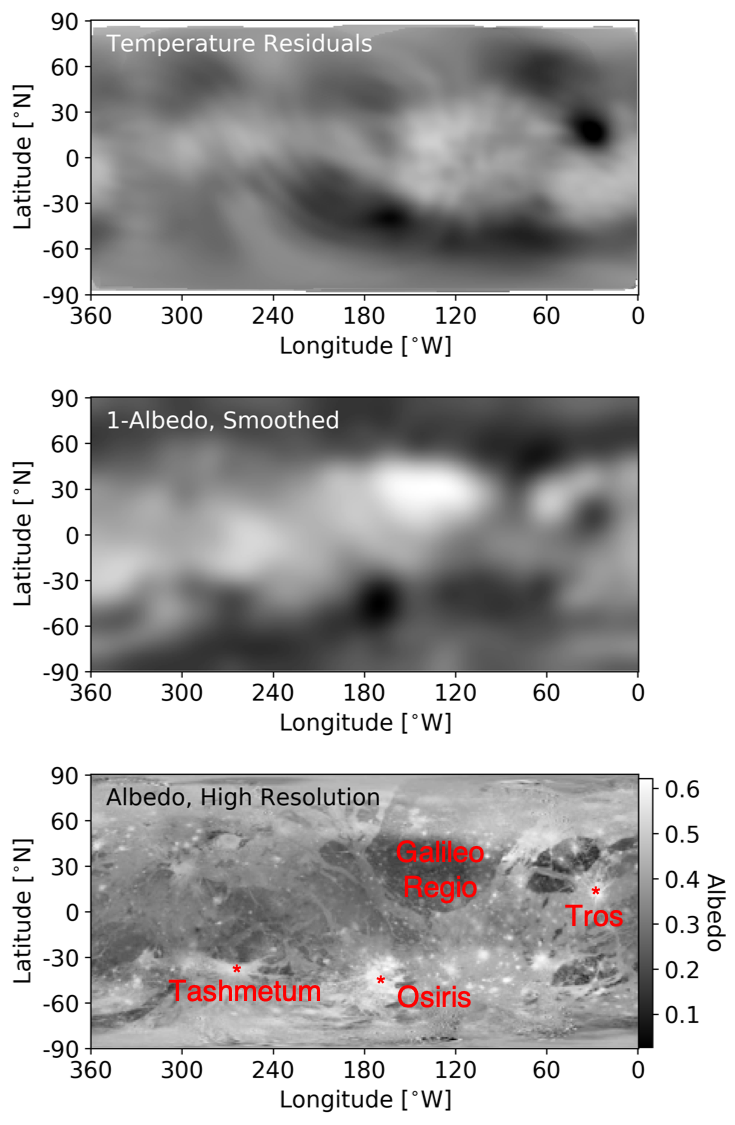}
\caption{Temperature residuals (data-model) from Figure \ref{fig:ResMap} shown alongside the albedo map (lower panel), and 1-Albedo, smoothed for comparison with the data. The albedo map is derived from \textit{Voyager} and \textit{Galileo} observations as described in Section \ref{sec:albedomap}. The thermal model assumes an albedo distribution based on spacecraft observations; incorrect albedos could therefore result in artifacts in derived thermal properties, which would be seen by residuals that track the albedo. While some albedo features correspond to thermal anomalies, there is not a systematic correspondence between temperature residuals and albedo, lending confidence to the treatment of albedo in the model. Even 10\% localized errors in the assumed albedos would fall short of explaining the magnitude of the observed temperature residuals. \label{fig:AlbedoComp}}
\end{figure}

\begin{figure}[ht]
\centering
\includegraphics[width=12cm]{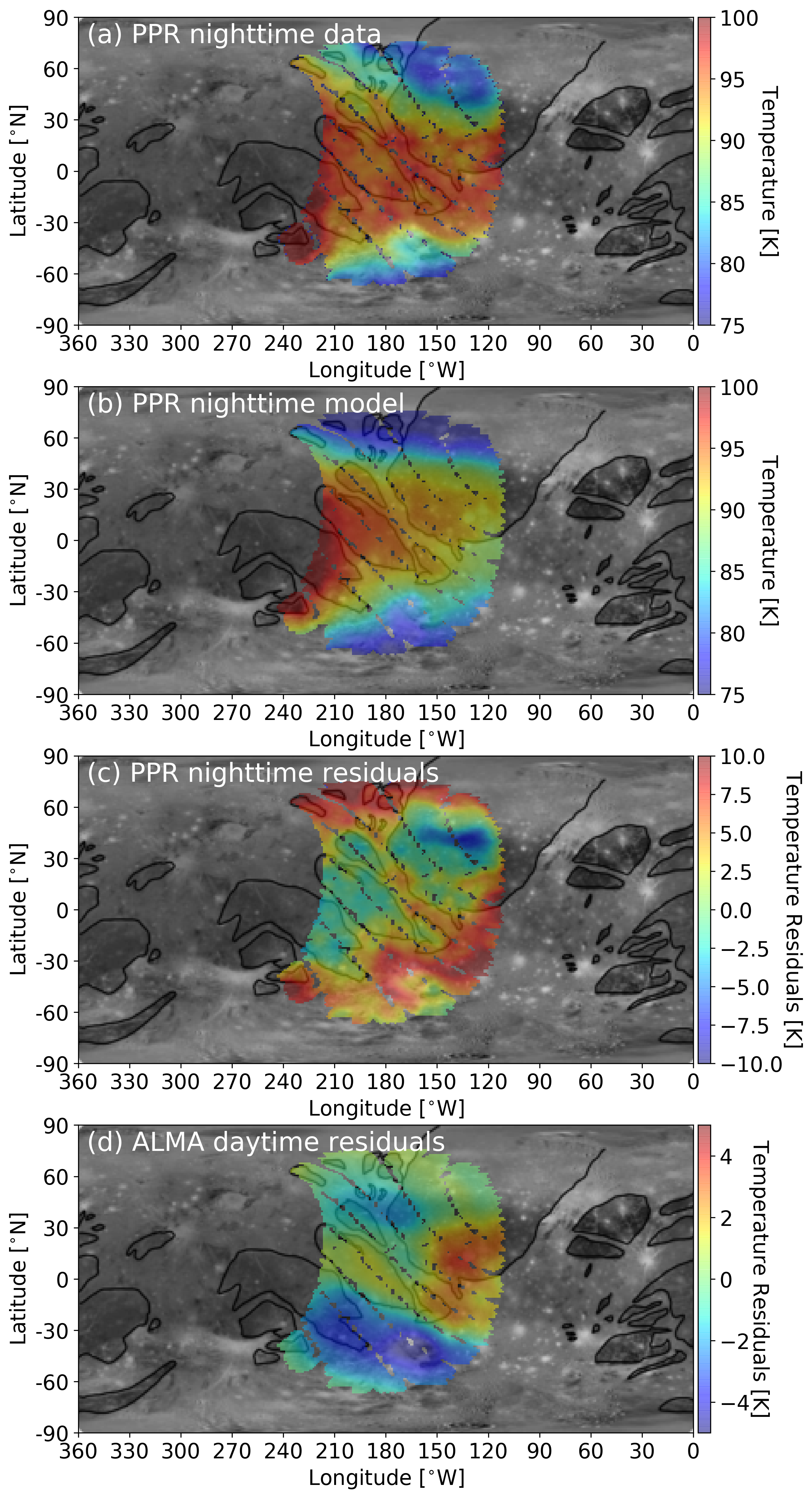}
\caption{\textit{Galileo} PPR observation of Ganymede at night in the $>$42 $\mu$m filter: (a) Temperature map; (b) Temperature model, assuming a high and low thermal inertia component homogeneously mixed across the surface; (c) Residuals after model subtraction; and (d) ALMA daytime temperature residual map as shown in Figure \ref{fig:ResMap}, masked to the same region. In all panels the surface temperatures for the region covered by PPR are shown superposed on a map of Ganymede's surface, with the dark terrains outlined in black. \label{fig:PPR}}
\end{figure}

Ganymede's bright impact craters Tros (27$^{\circ}$W, 11$^{\circ}$N) and Osiris (166$^{\circ}$W 38$^{\circ}$S; both labeled on Figure \ref{fig:AlbedoComp} for reference) appear 3-8 K colder than expected after accounting for their high albedos and for the temperature-dependent electrical skin depth (see Figures \ref{fig:Tmaps} \& \ref{fig:ResMap}). The bright impact crater Tros is a particularly striking thermal feature in the maps. This crater was observed in all three 233 GHz observations and can be seen even in the original images prior to model subtraction (Figure \ref{fig:data}); it is colder than the surrounding regions by 20-25 K, and colder than predicted by the model by 5-10 K. Note that a 10\% error in the assumed albedo could only account for about half of this deviation. The thermal anomaly associated with this crater is extended to the northwest from the center of the impact site; Figure \ref{fig:data} shows that the thermal feature is both larger than the spatial resolution element and oriented in a different direction from the beam in 2019, indicating that the temperature anomaly is both extended and resolved, and that its orientation is real rather than a reflection of the beam orientation. Moreover, the cold region is not centered over the crater, but is extended in the direction of greater ray extension.\par
While certainly among the brightest and freshest of Ganymede's craters, Tros exhibits an outsized thermal signature compared to its brightness and apparent size in optical images. It is also the only large, bright crater equatorward of the polar caps and falls directly in the area with maximum plasma influx (Figure \ref{fig:ResMap}). This region has the highest violet/green color ratio of any location on Ganymede's surface, which Khurana et al. (2007) use as a proxy for the presence of small-grain-size frost. Small grain sizes should increase the effective millimeter emissivity, contrary to the observed effect, but also increases the thermal inertia by increasing grain contact area, leading to colder daytime temperatures. In addition, the purity of the fresher ice in these terrains should increase the depth of subsurface sounding relative to their dirtier surroundings, accessing deeper (colder) temperatures as well as increasing the opportunities for volume scattering. Figure \ref{fig:Tvst} shows, for the example case of Tros, that the observed effective temperature can be matched by either a local increase in porosity or decrease in emissivity relative to the global model; the vertical temperature profiles are shown for different porosity profile cases in Figure \ref{fig:Tz}. Infrared data also find the impact craters to be colder during the day as well as at night than their surroundings, consistent with both their higher albedos and different regolith properties (see Section \ref{sec:ppr}).\par
Other anomalously cold regions in the thermal maps also align with collections of small, bright impact craters, although the association is less clear than in the case of Tros and Osiris. In particular, the cold region around 60-80$^{\circ}$W 60$^{\circ}$S coincides with a group of bright impact craters, but does not extend to the impact craters slightly north and west of this area. Similarly, the area near 300$^{\circ}$W 45$^{\circ}$N, as well as the region east of Tros extending to roughly 60$^{\circ}$W, are both somewhat cold relative to the model and coincide with clusters of impact craters, but other nearby clusters do not exhibit similar thermal effects. \par
Certain large impact features such as Tashmetum at 95$^{\circ}$W 39$^{\circ}$S (labeled on Figure \ref{fig:AlbedoComp} for reference) do not appear anomalous in thermal properties at all, despite being classified along with Tros and Osiris as fresh crater material by Collins et al. (2014). In general, only the brightest craters are associated with anomalous thermal features, consistent with surface aging processes that darken surfaces and/or lower the near-surface densities over time.\par
Craters with anomalously low brightness temperature are commonly observed on other icy satellites, including Pwyll crater on Europa (Trumbo et al. 2017); several craters on Titan (Janssen et al. 2016); and Inktomi on Rhea (Bonnefoy et al. 2020). The effect has been attributed to high thermal inertia in the case of Europa, and in the case of Rhea either reduced emissivity due to larger penetration depth  (Bonnefoy et al. 2020) or higher thermal inertia (Howett et al. 2014).
\subsection{Large-scale trends}\label{sec:res_trends}
The thermal maps (Figures \ref{fig:Tmaps} \& \ref{fig:ResMap}) show large-scale trends that are latitudinal and hemispherical in nature. Broadly, Ganymede is too warm in equatorial regions compared to the models by 3-5 K, and too cold between 30-60$^{\circ}$ in both the northern and southern hemisphere by a similar amount. This trend does not continue all the way to the poles: the polar regions, although poorly sensed by our Earth-based viewing geometry, are well fit by the model and if anything slightly warmer than predicted. The leading hemisphere equatorial region in particular appears too warm relative to its albedo, in a roughly elliptical area spanning 60-150$^{\circ}$W, while the trailing hemisphere exhibits a more compact too-warm region near its center (0$^{\circ}$N 270$^{\circ}$W).\par
Ganymede's poles are coated in water ice caps that extend down to roughly $\pm$45$^{\circ}$ in latitude, and may be the result of thermal mobilization and transport of ice from the warmer equatorial regions (Moore et al. 1996; Hillier et al. 1996). Although our results show a latitudinal trend in thermal properties, the brightness temperature is neither monotonically increasing nor decreasing towards higher latitudes relative to the thermal model. There is therefore no distinct thermal signature attributable to the polar caps themselves beyond what can be accounted for by their albedo and solar incidence angle, although they are incompletely covered by the data. \par
Instead, the correlation between the brightness temperature residuals and distance from the leading and trailing point is suggestive of exogenic modification due to bombardment by micrometeorites and neutral and ionized material. Bombardment by non-ionized material would preferentially affect the leading hemisphere (McKinnon \& Parmentier 1986), and would act to lower the bulk density and hence thermal inertia, allowing these regions to heat up faster in the daytime, which is consistent with the observed distribution. Bombardment also affects the effective emissivity in competing ways: an increase in porosity increases the transparency of the ice thus lowering the brightness temperature, while deposition of non-ice material decreases ice transparency. These effects cannot be compared quantitatively without better experimental data, but the observed brightness temperature trends imply that the balance between the processes favors those that lower thermal inertia and/or decrease transparency in equatorial regions. \par
Spatial variations in grain size arising from exogenic processes may also play a significant role through the effect of grain size on both emissivity and thermal conductivity. The low emissivity of icy satellites at millimeter and centimeter wavelengths is not fully understood, but has been attributed to surface roughness or volume scattering effects (e.g. Muhleman \& Berge 1991; Le Gall et al. 2017). In such a case, emissivity is minimized for a particle size of $\sim \lambda/(4\pi)$ (discussed in e.g. Lellouch et al. 2000; 2017). On Ganymede, the near-infrared ice bands have been used to infer a particle size decreasing from around 1 mm in equatorial regions to tens of $\mu$m at the poles (Ligier et al. 2019; Stephan et al. 2020). If emissivity variations are indeed controlled by scattering, this would imply an emissivity minimum at mid-latitudes, which matches the observed brightness temperature residuals qualitatively although the corresponding dependence of the latitude of minimum emissivity on frequency is not observed. Thermal conductivity and hence thermal inertia is also higher for smaller grains (see Figure \ref{fig:kcond}); the effect of decreasing grain size with latitude would therefore be higher thermal inertias in more polar regions. \par
Ionized material impacts Ganymede's surface at mid-latitudes and above; equatorial regions are protected from the jovian magnetosphere by Ganymede's intrinsic magnetic field (Khurana et al. 2007; Poppe et al. 2018; McGrath et al. 2013). Incident material affects the surface density and conductivity in competing ways through the effects of sputtering and sintering, and decreases the transparency of the ice by depositing non-ice material. Mimas' cold leading-hemisphere daytime temperatures have been attributed to crystalline ice sintered by high-energy electron bombardment, which increases thermal inertia (Howett et al. 2011; Ferrari \& Lucas 2016). On Ganymede, the colder daytime regions coincide with the mid-latitudes where the electric field lines reach down to the surface; this correlation is consistent with a similar effect at work, although the actual regions impacted by incoming plasma (outlines shown in Figure \ref{fig:ResMap}) do not align well with the areas in our observations that are coldest relative to the models.\par
Thus while several exogenic processes are consistent with aspects of the spatial trends in thermal properties, no single mechanism can fully explain the observed trends and it is likely that the thermophysical properties of Ganymede's near-surface are controlled by a combination of the mechanisms described above rather than dominated by one particular process.
\section{Conclusions}\label{sec:conc}
We present spatially-resolved thermal continuum observations of Ganymede at millimeter wavelengths covering all longitudes and three frequencies sensitive to the upper cm through $\sim$0.5 m of the surface. Our global best-fit models have spectral emissivities of 0.75-0.78, and porosities of 40$\pm30$\% and 10$^{+30}_{-10}$\%, corresponding to effective thermal inertias of 400$^{+300}_{-200}$ J m$^{-2}$ K$^{-1}$ s$^{-1/2}$ at 233 and 343.5 GHz (1.3 and 0.87 mm) and of 750$^{+200}_{-350}$ J m$^{-2}$ K$^{-1}$ s$^{-1/2}$ at 97.5 GHz (3 mm) respectively. These thermal inertias are higher than previously measured for Ganymede in the infrared, even taking into account the large uncertainties, but the infrared and millimeter data are collectively consistent with a compaction gradient from a surface porosity of 85\% to a deep porosity of 10\% (and no greater than 40\% considering uncertainties), over a compaction length scale in the tens of cm range. The lower millimeter emissivities compared to the infrared may be attributable to surface or volume scattering, which will preferentially affect millimeter radiation given the particle sizes of 50 $\mu$m - 1 mm (Stephan et al. 2020), and/or to an underestimation of electrical skin depth in our model due to unknowns in the surface composition and regolith particle properties.\par
The global best-fit properties can match the observed temperature distribution across the surface to within 10 K. Localized deviations from the model temperatures are well above the noise level and are consistent between observations at different times, viewing geometries, and frequencies, indicative of robust regional-scale variations in thermal surface properties. We find that some dark terrains are associated with positive thermal anomalies that require both an increased porosity (lower thermal inertia) and an increased dust fraction (lower transparency) to match. However, the correlation between thermal properties and terrain type is incomplete and not present at all surface locations, in contrast to what is seen in \textit{Galileo} PPR data. In addition, at millimeter wavelengths the equatorial regions are warmer than predicted by the model by about 3-5 K, in particular near the centers of the leading and trailing hemispheres, while the mid-latitudes (roughly 30-60$^{\circ}$) are colder than predicted by 3-5 K in both hemispheres. These trends, as well as localized temperature deviations, are broadly consistent across the three ALMA frequency bands, despite differing from the infrared data.\par
The bright impact craters, most prominently Tros and Osiris, are colder than expected by 5-8 K even after accounting for albedo and for temperature-dependent dielectric properties, which lower the brightness temperature by a few K due to the high transparency of cold porous ice at 100-350 GHz. Optical and UV data have indicated that these bright impact terrains contain a high water ice abundance and small ice grains (Khurana et al. 2007), the former of which would increase the transparency of the ice to millimeter wavelengths and the latter of which would have the reverse effect on transparency, but would increase thermal inertia. \par
ALMA observations of Europa likewise found that Pwyll crater exhibits a higher thermal inertia than surrounding regions (Trumbo et al. 2017), and Rathbun \& Spencer (2020) subsequently obtained the same result using \textit{Galileo} PPR infrared data. An additional anomalously cold region on Europa was identified with ALMA (Trumbo et al. 2018); although this region does not correspond to an impact crater, it does similarly align with the area of greatest water ice surface abundance from near-infrared spectroscopy (Brown \& Hand 2013). \par
On Ganymede, the fact that the large-scale trends in thermal features are largely symmetric across the equator and between hemispheres, and in particular that there is a correlation with distance from the leading and trailing points, suggests an exogenic origin. Bombardment by micrometeorites and neutral material would preferentially affect the leading and trailing hemispheres, acting to increase porosity, add non-ice contaminants to the surface, and perhaps alter grain size. These processes would have differing effects on the observed brightness temperature; the fact that the leading and trailing points are warmer than expected suggests that either ice contamination increasing the emissivity, or a decreased thermal inertia due to an increased porosity, dominate over the effect of increased porosity on depressing emissivity. \par
Plasma bombardment at and above mid-latitudes results in the competing effects of sputtering and sintering, which affect both porosity and grain size and moreover vary with latitude as the penetration depth of electrons and ions is energy-dependent. Our observations show that the mid-latitudes are colder than expected, but not the polar regions. The many competing effects that vary with latitude (grain size, temperature-dependent thermal conductivity and dielectric properties, and compositional gradients due to thermal segregation, in addition to energy of incident particles) make it challenging to conclusively attribute this trend to a specific process. It is notable that the gradient in grain size from equator to pole (Stephan et al. 2020) would favor an emissivity minimum at exactly these latitudes, although this interpretation is challenged by the lack of frequency dependence in the latitudinal trend that should be present if grain size dependent scattering were responsible. \par
The data presented here represent the first spatially and vertically resolved observations of Ganymede's surface at these depths, and reveal thermal signatures that don't correlate to any known compositional or geological units. Future lower-frequency observations that are sensitive to deeper in the surface, as well as polarization measurements, would put new constraints on model parameters and hence reduce degeneracies in interpretation. Interpretation of data at these frequencies would be greatly aided by experimental data on the dielectric properties of ice as a function of density, grain size, and impurities, at frequencies of 100-500 GHz and temperatures of 100-200 K relevant for the Galilean satellites. The higher spatial resolutions that are now possible with ALMA's long baselines configurations will also reveal the distribution of Ganymede's surface properties in greater detail, while the JUICE Submillimetre Wave Instrument (SWI) will extend this work to higher frequencies ($>$500 GHz) and hence help to close the gap between millimeter and infrared depth sensitivities. Such data may provide the added information to independently constrain a greater number of surface properties and unravel how the range of exogenic processes acting on Ganymede's surface together produce the distributions presented here.
\clearpage

\clearpage
\section*{Acknowledgements}
This work was supported in part by the Heising-Simons Foundation \textit{51 Pegasi b} postdoctoral fellowship to K. de Kleer. This paper makes use of the following ALMA data: ADS/JAO.ALMA\#2016.1.00691.S, ADS/JAO.ALMA\#2018.1.01292.S, ADS/JAO.ALMA\#2011.0.00001.CAL. ALMA is a partnership of ESO (representing its member states), NSF (USA) and NINS (Japan), together with NRC (Canada), MOST and ASIAA (Taiwan), and KASI (Republic of Korea), in cooperation with the Republic of Chile. The Joint ALMA Observatory is operated by ESO, AUI/NRAO and NAOJ. The National Radio Astronomy Observatory is a facility of the National Science Foundation operated under cooperative agreement by Associated Universities, Inc.
\clearpage
\section*{References}
\begin{itemize}[label={}]
\item Altenhoff, W.J., Chini, R., Hein, H., et al. 1988. First radio astronomical estimate of the temperature of Pluto. \textit{A\&A Letter} \textbf{190}. L15-L17.
\item Andrae, R. 2010. Error estimation in astronomy: A guide. arXiv:1009.2755v3.
\item Bonnefoy, L.E., Le Gall, A., Lellouch, E., et al. 2020. Rhea's subsurface probed by the Cassini radiometer: Insights into its thermal, structural, and compositional properties. \textit{Icarus} 352, article id. 113947.
\item Bottke, W.F., Vokrouhlick\'y, D., Nesvorn\'y, D., Moore, J.M. 2013. Black rain: The burial of the Galilean satellites in irregular satellite debris. \textit{Icarus} \textbf{223}, 775-795.
\item Brouet, Y., Neves, L., Sabouroux, P., et al. 2016. Characterization of the permittivity of controlled porous water ice-dust mixtures to support the radar exploration of icy bodies. JGR Planets 121, 2426-2443.
\item Brown, M.E. \& Hand, K.P. 2013. Salts and radiation products on the surface of Europa. AJ 145:110, 7pp.
\item Buratti, B.J. 1991. Ganymede and Callisto: Surface textural dichotomies and photometric analysis. Icarus \textbf{92}, 312-323.
\item Butler, B. 2012. Flux density models for Solar System bodies in CASA. ALMA Memo \# 594.
\item Choy, T.C. 1999. Effective medium theory. Oxford: Clarendon Press.
\item Clark, R.N. \& McCord, T.B. 1980. The Galilean satellites: New near-infrared spectral reflectance (0.65-2.5 microns) and a 0.325-0.5 micron summary. Icarus \textbf{41}, 323-329.
\item Collins, G.C., Patterson, G.W., Head, J.W., et al. 2014. Global geologic map of Ganymede: U.S. Geological Survey Scientific Investigations Map 3237.
\item Cornwell, T. \& Fomalont, E.B. 1999. Self-Calibration. In: \textit{Synthesis Imaging in Radio Astronomy II} ed. Taylor, Carilli, and Perley. \textit{Astronomical Society of the Pacific Conference Series} \textbf{180}, 187pp.
\item de Pater, I., Brown, R.A., \& Dickel, J.R. 1984. VLA observations of the Galilean satellites. \textit{Icarus} \textbf{57}, 93-101.
\item de Pater, I., Luszcz-Cook, S., Rojo, P., et al. 2020. ALMA observations of Io going into and coming out of eclipse. \textit{PSJ} 1:60, 25pp.
\item Ennis, D.J., Neugebauer, G. \& Werner, M. 1982. Millimeter continuum observations of quasars. ApJ 262, 460-477.
\item Ferrari, C. \& Lucas, A. 2016. Low thermal inertias of icy planetary surfaces: Evidence for amorphous ice? \textit{A\&A} \textbf{588}, A133, 14pp.
\item Foreman-Mackey, D., Hogg, D.W., Lang, D. \& Goodman, J. 2013. emcee: The MCMC hammer.  arXiv:1202.3665v4.
\item Francis, L., Johnstone, D., Herczeg, G., Hunter, T.R. \& Harsono, D. 2020. On the accuracy of the ALMA flux calibration in the time domain and across spectral windows. arXiv:2010.02186.
\item Goodman, J. \& Weare, J. 2010. Ensemble samplers with affine invariance. Comm. Appl. Math. and Comp. Sci, \textbf{5}, 65-80.
\item Gundlach, B. \& Blum, J. 2012. Outgassing of icy bodies in the Solar System - II: Heat transport in dry, porous surface dust layers. Icarus 219, 618-629.
\item Gusarov, A.V., Laoui, T., Froyen, L. \& Titov, V.I. 2003. Contact thermal conductivity of a powder bed in selective laser sintering. Int. J. Heat Mass Transfer 46, 1103-1109.
\item Hanel, R., Conrath, B., Flasar, M. et al. 1979. Infrared observations of the jovian system from Voyager 2. \textit{Science} \textbf{206}, 952-956.
\item Hansen, G.B. \& McCord, T.B. 2004. Amorphous and crystalline ice on the Galilean satellites: A balance between thermal and radiolytic processes. \textit{JGR} \textbf{109}, CiteID E01012.
\item Hayne, P.O., Bandfield, J.L., Siegler, M.A. et al. 2017. Global regolith thermophysical properties of the Moon from the Diviner Lunar Radiometer Experiment. JGR: Planets 122, 2371-2400.
\item Heggy, E., Palmer, E.M., Kofman, W. et al. 2012. Radar properties of comets: Parametric dielectric modeling of Comet 67P/Churyumov-Gerasimenko. Icarus 221, 925-939.
\item Hewison, T.J. \& English, S. 1999. Airborne retrievals of snow and ice surface emissivity at millimeter wavelengths. IEEE Trans. on Geo. and Rem. Sens. 37, 1871-1879.
\item Hillier, J., Helfenstein, P. \& Veverka, J. 1996. Latitude variations of the polar caps on Ganymede. \textit{Icarus} \textbf{124} 308-317.
\item Howett, C.J.A., Spencer, J.R., Schenk, P., et al. 2011. A high-amplitude thermal inertia anomaly of probable magnetospheric origin on Saturn's moon Mimas. \textit{Icarus} \textbf{216}, 221-226.
\item Howett, C.J.A., Spencer, J.R., Hurford, T., Verbiscer, A. \& Segura, A. 2014. Thermophysical property variations across Dione and Rhea. Icarus 241, 239-247.
\item Hufford, G. 1991. A model for the complex permittivity of ice at frequencies below 1 THz. Int. J. Infrared Millimeter Waves 12, 677-682.
\item Janssen, M.A., Le Gall, A., Lopes, R.M., et al. 2016. Titan's surface at 2.18-cm wavelength imaged by Cassini RADAR radiometer: Results and interpretations through the first ten years of observation. Icarus 270, 443-459.
\item Jia, X., Walker, R.J., Kivelson, M.G., Khurana, K.K. \& Linker, J.A. 2008. Three-dimensional MHD simulations of Ganymede's magnetosphere. JGR: Space Physics 113, id.A05212, 17pp.
\item Jia, X., Walker, R.J., Kivelson, M.G., Khurana, K.K. \& Linker, J.A. 2009. Properties of Ganymede's magnetosphere inferred from improved three-dimensional MHD simulations. JGR: Space Physics 114, id.A09209, 19pp.
\item Jiang, J.H. \& Wu, D.L. 2004. Ice and water permittivities for millimeter and sub-millimeter remote sensing applications. Atmospheric Science Letters 5, 146-151.
\item Johnson, K.L., Kendall, K. \& Roberts, A.D. 1971. Surface energy and the contact of elastic solids. Proc. of the Roy. Soc. of London 324, 301-313.
\item Johnson, T.V., Soderblom, L.A., Mosher, J.A., et al. 1983. Global multispectral mosaics of the icy galilean satellites. JGR \textbf{88}, 5789-5805.
\item Khurana, K.K., Pappalardo, R.T., Murphy, N. \& Denk, T. 2007. The origin of Ganymde's polar caps. Icarus 191, 193-202.
\item Kivelson, M.G., Khurana, K.K. \& Volwerk, K. 2002. The permanent and inductive magnetic moments of Ganymede. \textit{Icarus} \textbf{157}, 507-522.
\item Klinger, J. 1980. Influence of a phase transition of ice on the heat and mass balance of comets. Science 209, 271-272.
\item Ligier, N., Paranicas, C., Carter, J. et al. 2019. Surface composition and properties of Ganymede: Updates from ground-based observations with the near-infrared imaging spectrometer SINFONI/VLT/ESO. \textit{Icarus} \textbf{333}, 496-515.
\item Le Gall, A., Leyrat, C., Janssen, M.A., et al. 2014. Iapetus' near surface thermal emission modeled and constrained using Cassini RADAR Radiometer microwave observations. Icarus 241, 221-238.
\item Le Gall, A., Leyrat, C., Janssen, M.A. et al. 2017. Thermally anomalous features in the subsurface of Enceladus' south polar terrain. Nature Asronomy 1, id. 0063.
\item Lellouch, E., Paubert, G., Moreno, R. \& Schmitt, B. 2000. NOTE: Search for variations in Pluto's millimeter-wave emission. Icarus 147, 580-584.
\item Lellouch E., Moreno, R., M\:uller, T., et al. 2017. The thermal emission from Centaurs and trans-Neptunian objects at millimeter wavelengths from ALMA observations. A\&A 608, A45, 21pp.
\item McCord, T.B., Carlson, R., Smythe, W., et al. 1997. Science 278, 271-275.
\item McGrath, M.A., Jia, X., Retherford, K. et al. 2013. Aurora on Ganymede. JGR Space Physics 118, 2043-2054.
\item McKinnon, W.B. \& Parmentier, E.M. 1986. Ganymede and Callisto. In: Satellites, University of Arizona Press, p. 718-763.
\item McMullin, J.P., Waters, B., Schiebel, D., Young, W. \& Golap, K. 2007. CASA Architecture and Applications. Astronomical Data Analysis Software and Systems XVI ASP Conference Series, Vol. 376, proceedings of the conference held 15-18 October 2006 in Tucson, Arizona, USA. Edited by Richard A. Shaw, Frank Hill and David J. Bell., p.127.
\item Mishima, O., Klug, D.D. \& Wahlley, E. 1983. The far-infrared spectrum of ice Ih in the range 8-25 cm$^{-1}$. Sound waves and difference bands, with application to Saturn's rings. \textit{J. Chem. Phys.} 78, 6399-6404.
\item Mitchell, D.L. \& de Pater, I. 1994. Microwave imaging of Mercury's thermal emission at wavelengths from 0.3 to 20.5 cm. \textit{Icarus} \textbf{110}, 2-32. 
\item Moore, J.M. , Mellon, M.T. \& Zent, A.P. 1996. Mass wasting and ground collapse in terrains of volatile-rich deposits as a solar system-wide geological process: The pre-\textit{Galileo} view. \textit{Icarus} \textbf{122}, 63-78.
\item Moreno, R. 2007. Report on continuum measurements of Ganymede and Callisto with the IRAM-PdB interferometer: Application to flux calibration, Internal Memo.
\item Morrison, D. 1969. Thermal models and microwave temperatures of the planet mercury. \textit{SAO Special Report} \#292.
\item Morrison, D. \& Cruikshank, D.P. 1973. Thermal properties of the Galilean satellites. \textit{Icarus} \textbf{18}, 224-236.
\item Muhleman, D.O., Berge, G.L., Rudy, D.J. \& Niell, A. 1986. Precise VLA positions and flux density measurements of the Jupiter System. AJ 92, 1428-1435.
\item Muhleman, D.O. \& Berge, G.L. 1991. Observations of Mars, Uranus, Neptune, Io, Europa, Ganymede, and Callisto at a wavelength of 2.66 mm. \textit{Icarus} \textbf{92}, 263-272.
\item Nelder, J.A. \& Mead, R. 1965. A simplex method for function minimization. \textit{Computer Journal} \textbf{7}, 308-313.
\item Ostro, S.J. 1982. Radar properties of Europa, Ganymede, and Callisto. In \textit{Satellites of Jupiter} (D. Morrison, Ed.)., Univ. of Arizona Press, Tucson.
\item Pappalardo, R.T., Collins, G.C. \& Head, J.W. 2004. Geology of Ganymede. In: \textit{Jupiter: The planet, satellites and magnetosphere}, ed. F. Bagenal, T. Dowling, W. McKinnon. Cambridge University Press, New York. pp. 363-396.
\item Poppe, A., Fatemi, S. \& Khurana, K. 2018. Thermal and energetic ion dynamics in Ganymede's magnetosphere. JGR: Space Physics 123, 4613-4637.
\item Pauliny-Toth, I.I.K., Witzel, A. \& Gorgolewski, S. 1997. Observations of Ganymede and Callisto at 1.3 cm wavelength. A\&A 58, L27-L28.
\item Prockter, L.M., Figueredo, P.H., Pappalardo, R.T., Head, J.W., Collins, G.C. 2000. JGR 105, 22519-22540.
\item Queipo, N.V., Haftka, R.T., Shyy, W., et al. 2005. Surrogate-based analysis and optimization. \textit{Progress in Aerospace Sciences} \textbf{41}, 1-28.
\item Rathbun, J.A. \& Spencer, J.R. 2020. Proposed plume source region on Europa: No evidence for endogenic thermal emission. \textit{Icarus} 338, article id. 113500.
\item Rau, U. \& Cornwell, T.J. 2011. A multi-scale multi-frequency deconvolution algorithm for synthesis imaging in radio interferometry. \textit{A\&A} \textbf{532}, id. A71, 17pp.
\item Rudy, D.J., Muhleman, D.O., Berge, G.L., Jakosky, B.M. \& Christensen, P.R. 1987. Mars: VLA observations of the northern hemisphere and the north polar region at wavelengths of 2 and 6 cm. \textit{Icarus} \textbf{71}, 159-177.
\item Russell, E.E., Brown, F.G., Chandos, R.A. et al. 1992. Galileo photopolarimeter/radiometer experiment. \textit{Space Sci. Rev.} \textbf{60}, 531-563.
\item Schenk, P.M., McKinnon, W.B., Gwynn, D., Moore, J.M. 2001. Nature 410, 57-60.
\item Shulman, L.M. 2004. The heat capacity of water ice in interstellar or interplanetary conditions. A\&A 416, 187-190.
\item Spencer, J.R. 1987. The surfaces of Europa, Ganymede, and Callisto: An investigation using Voyager IRIS thermal infrared spectra. PhD dissertation, Lunary and Planet. Lab., Univ. of Ariz., Tucson.
\item Spencer, J.R., Lebofsky, L.A. \& Sykes, M.A. 1989. Systematic biases in radiometric diameter determinations. \textit{Icarus} \textbf{78}, 337-354.
\item Squyres, S.W. \& Veverka, J. 1982. Variation of albedo with solar incidence angle on planetary surfaces. Icarus 50, 112-122.
\item Stephan, K., Hibbits, C.A. \& Jaumann, R. 2020. H$_2$O-ice particle size variations across Ganymede's and Callisto's surface. \textit{Icarus} \textbf{337}, 113440, 21pp.
\item Tikhonova, T.V. \& Troitski, V.S. 1969. Effect of heat from within the moon on its radio emission for the case of lunar properties which vary with depth. \textit{Soviet Astronomy} \textbf{13}, 120-128.
\item Tiuri, M.E., Sihvola, A.H., Nyfors, E.G. \& Hallikaiken, M.T. 1984. The complex dielectric constant of snow at microwave frequencies. IEEE Journal of Oceanic Engineering 9, 377-382.
\item Trumbo, S.K., Brown, M.E. \& Butler, B.J. 2017. ALMA thermal observations of a proposed plume source region on Europa. AJ 154, article id. 148, 6 pp.
\item Trumbo, S.K., Brown, M.E. \& Butler, B.J. 2018. ALMA thermal observations of Europa. AJ 156, article id. 161, 7 pp.
\item Ulich, B.L. \& Conklin, E.K. 1976. Observations of Ganymede, Callisto, Ceres, Uranus, and Neptune. \textit{Icarus} \textbf{27}, 187-189.
\item Ulich, B.L. 1981. Millimeter-wavelength continuum calibration sources. \textit{AJ} \textbf{86}, 1619-1626.
\item Ulich, B.L., Dickel, J.R. \& de Pater, I. 1984. Planetary observations at a wavelength of 1.32 mm. \textit{Icarus} \textbf{60}, 590-598.
\item Yan, B., Weng, F. \& Meng, H. 2008. Retrievals of snow surface microwave emissivity from the advanced microwave sounding unit. JGR: Atmospheres 113, D19206, 23pp.
\item Young, L.A. 2017. Volatile transport on inhomogeneous surfaces: II. Numerical calculations (VT3D). Icarus 284, 443-476.

\end{itemize}

\end{document}